\documentclass[longauth]{aa} 
\usepackage{graphicx}
\usepackage{color}

\usepackage[varg]{txfonts}
\usepackage{lscape,supertabular}
\usepackage{arydshln}
\newcommand{\mearth}{M_\oplus}
\newcommand{\rearth}{R_{\rm \oplus}}
\newcommand{\msun}{M_\odot}
\newcommand{\rsun}{R_{\rm \odot}}
\newcommand{\lsun}{L_\odot}
\newcommand{\rhosun}{\rho_\odot}

\newcommand{\logg}{\log g}
\newcommand{\feh}{[\mathrm{Fe}/\mathrm{H}]}

\def\ms{\hbox{\,m\,s$^{-1}$}}         
\def\m2s2{\hbox{\,m$^{2}$\,s$^{-2}$}} 
\def\sini{\hbox{sin\,$i$}}      

\begin{document}

   \title{Eyes on K2-3: A system of three likely sub-Neptunes characterized with HARPS-N and HARPS}


   \author{M.~Damasso\inst{\ref{oato}}
   \and A. S.~Bonomo\inst{\ref{oato}}       
   \and N.~Astudillo-Defru \inst{\ref{obsge}}
   \and X.~Bonfils \inst{\ref{ipag}}
   \and L.~Malavolta\inst{\ref{pduni},\ref{oapd}}
   \and A.~Sozzetti\inst{\ref{oato}}
   \and E.~Lopez\inst{\ref{godd}}
   \and L.~Zeng\inst{\ref{harvplan}}
   \and R. D.~Haywood\inst{\ref{cfa}}
   \and J. M.~Irwin\inst{\ref{cfa}}
   \and A.~Mortier \inst{\ref{supa}}
   \and A.~Vanderburg\inst{\ref{unitx},\ref{cfa},\ref{sagan}}
   \and J.~Maldonado\inst{\ref{oapa}}
   \and A. F.~Lanza\inst{\ref{oact}}
   \and L.~Affer\inst{\ref{oapa}}
   \and J.-M.~Almenara \inst{\ref{obsge}}
   \and S.~Benatti\inst{\ref{oapd}}
   \and K.~Biazzo\inst{\ref{oact}}
   \and A.~Bignamini\inst{\ref{oats}}
   \and F.~Borsa\inst{\ref{brera}}
   \and F.~Bouchy \inst{\ref{obsge}}
   \and L. A.~Buchhave\inst{\ref{bohr}}
   \and A. C.~Cameron\inst{\ref{supa}}
   \and I.~Carleo\inst{\ref{oapd}}
   \and D.~Charbonneau\inst{\ref{cfa}}
   \and R.~Claudi\inst{\ref{oapd}}
   \and R.~Cosentino\inst{\ref{tng}}
   \and E.~Covino\inst{\ref{oana}}
   \and X.~Delfosse \inst{\ref{ipag}}
   \and S.~Desidera\inst{\ref{oapd}}
   \and L.~Di Fabrizio\inst{\ref{tng}}
   \and C.~Dressing\inst{\ref{berk}}
   \and M.~Esposito\inst{\ref{oana}}
   \and R.~Fares\inst{\ref{oact}}
   \and P.~Figueira\inst{\ref{esochile},\ref{uporto}}
   \and A. F. M.~Fiorenzano\inst{\ref{tng}}
   \and T.~Forveille \inst{\ref{ipag}}
   \and P.~Giacobbe\inst{\ref{oato}}
   \and E.~Gonz\'{a}lez-\'{A}lvarez\inst{\ref{oapa}}
   \and R.~Gratton\inst{\ref{oapd}}
   \and A.~Harutyunyan\inst{\ref{tng}}
   \and J.~Asher Johnson\inst{\ref{cfa}}
   \and D. W.~Latham\inst{\ref{cfa}}
   \and G.~Leto\inst{\ref{oact}}
   \and M. Lopez-Morales\inst{\ref{cfa}}
   \and C.~Lovis \inst{\ref{obsge}}
   \and A.~Maggio\inst{\ref{oapa}}
   \and L.~Mancini\inst{\ref{torver},\ref{mpi},\ref{oato}}
   \and S.~Masiero\inst{\ref{oapa}}
   \and M.~Mayor \inst{\ref{obsge}}
   \and G.~Micela\inst{\ref{oapa}}
   \and E.~Molinari\inst{\ref{tng},\ref{oacg}}
   \and F.~Motalebi\inst{\ref{obsge}}
   \and F.~Murgas \inst{\ref{iac},\ref{ull}}
   \and V.~Nascimbeni\inst{\ref{pduni},\ref{oapd}}
   \and I.~Pagano\inst{\ref{oact}}
   \and F.~Pepe \inst{\ref{obsge}}
   \and D. F.~Phillips\inst{\ref{cfa}}
   \and G.~Piotto\inst{\ref{pduni},\ref{oapd}}
   \and E.~Poretti\inst{\ref{brera}}
   \and M.~Rainer\inst{\ref{brera}}
   \and K.~Rice\inst{\ref{supa2}}
   \and N.~C.~Santos \inst{\ref{uporto},\ref{caup}}
   \and D.~Sasselov\inst{\ref{cfa}}
   \and G.~Scandariato\inst{\ref{oact}}
   \and D.~S\'egransan \inst{\ref{obsge}}
   \and R.~Smareglia\inst{\ref{oats}}
   \and S.~Udry \inst{\ref{obsge}}
   \and C.~Watson\inst{\ref{unibel}}
   \and A.~W\"unsche \inst{\ref{ipag}}
}
   \institute{INAF - Osservatorio Astrofisico di Torino, Via Osservatorio 20, I-10025 Pino Torinese, Italy\label{oato}\\
         \email{damasso@oato.inaf.it}         
  \and Observatoire de Gen\`eve, Universit\'e de Gen\`eve, 51 ch. des
Maillettes, 1290 Sauverny, Switzerland\label{obsge}
\and Univ. Grenoble Alpes, CNRS, IPAG, 38000 Grenoble,
France\label{ipag}
\and Dipartimento di Fisica e Astronomia ``Galileo Galilei", Universita'di Padova, Vicolo dell'Osservatorio 3, 35122 Padova, Italy\label{pduni}
\and INAF - Osservatorio Astronomico di Padova, Vicolo dell'Osservatorio 5, 35122 Padova, Italy\label{oapd}
\and NASA Goddard Space Flight Center, 8800 Greenbelt Rd, Greenbelt, MD 20771, USA \label{godd}
\and Department of Earth and Planetary Sciences, Harvard University, Cambridge, MA 02138 \label{harvplan}
\and Harvard-Smithsonian Center for Astrophysics, 60 Garden Street, Cambridge, MA 02138, USA\label{cfa}
\and Centre for Exoplanet Science, SUPA, School of Physics and Astronomy, University of St Andrews, St Andrews KY16 9SS, UK\label{supa}
\and Department of Astronomy, The University of Texas at Austin, 2515 Speedway, Stop C1400, Austin, TX 78712\label{unitx}
\and NASA Sagan Fellow\label{sagan}
\and INAF – Osservatorio Astronomico di Palermo, Piazza del Parlamento 1, I-90134 Palermo \label{oapa}
\and INAF – Osservatorio Astrofisico di Catania, Via S. Sofia 78, I-95123 Catania \label{oact}
\and INAF – Osservatorio Astronomico di Trieste, Via Tiepolo 11, I-34143 Trieste \label{oats}
\and INAF-Osservatorio Astronomico di Brera, Via E. Bianchi 46, 23807 Merate, Italy \label{brera}
\and Centre for Star and Planet Formation, Niels Bohr Institute \& Natural History Museum, University of Copenhagen, DK-1350 Copenhagen, Denmark \label{bohr}
\and INAF - Fundaci\'on Galileo Galilei, Rambla Jos\'e Ana Fernandez P\'erez 7, 38712 Bre\~na Baja, Spain \label{tng}
\and INAF – Osservatorio Astronomico di Capodimonte, Salita Moiariello 16, I-80131 Napoli \label{oana}
\and Department of Astronomy, University of California, Berkeley, CA 94720, USA \label{berk}
\and European Southern Observatory, Alonso de Cordova 3107, Vitacura, Santiago, Chile \label{esochile} 
\and Instituto de Astrof\'isica e Ci\^encias do Espa\c{c}o, Universidade
do Porto, CAUP, Rua das Estrelas, 4150-762 Porto, Portugal\label{uporto}
\and Dipartimento di Fisica, Universit`a di Roma Tor Vergata, Via della Ricerca Scientifica 1, 00133 – Roma, Italy \label{torver}
\and Max Planck Institute for Astronomy, Königstuhl 17, 69117 – Heidelberg, Germany \label{mpi}
\and INAF - Osservatorio di Cagliari, via della Scienza 5, 09047 Selargius, CA,
Italy \label{oacg}
\and Instituto de Astrof\'sica de Canarias (IAC), E-38200 La Laguna,
Tenerife, Spain\label{iac}
\and Dept. Astrof\'isica, Universidad de La Laguna (ULL), E-38206 La
Laguna, Tenerife, Spain\label{ull}
\and SUPA, Institute for Astronomy, University of Edinburgh, Royal Observatory, Blackford Hill, Edinburgh, EH93HJ, UK \label{supa2}
\and Departamento de F\'isica e Astronomia, Faculdade de Ci\^encias,
Universidade do Porto, Rua do Campo Alegre, 4169-007 Porto,
Portugal\label{caup}
\and Astrophysics Research Centre, School of Mathematics and Physics, Queen's University Belfast, Belfast BT7 1NN, UK \label{unibel}
}

   \date{Accepted February 20, 2018}

 
  \abstract
   {M-dwarf stars are promising targets for identifying and characterizing potentially habitable planets. K2-3 is a nearby (45 pc), early-type M dwarf hosting three small transiting planets, the outermost of which orbits close to the inner edge of the stellar (optimistic) habitable zone. The K2-3 system is well suited for follow-up characterization studies aimed at determining accurate masses and bulk densities of the three planets.}
   {Using a total of 329 radial velocity measurements collected over 2.5 years with the 
   HARPS-N and HARPS spectrographs and a proper treatment of the stellar activity signal, we aim to improve measurements of the masses and bulk densities of the K2-3 planets. We use our results to investigate the physical structure of the planets.}
   {We analyse radial velocity time series extracted with two independent pipelines by using Gaussian process regression. We adopt a quasi-periodic kernel to model the stellar magnetic activity jointly with the planetary signals. We use Monte Carlo simulations to investigate the robustness of our mass measurements of K2-3\,c and K2-3\,d, and to explore how additional high-cadence radial velocity observations might improve them.}
   {Despite the stellar activity component being the strongest signal present in the radial velocity time series, we are able to derive masses for both planet b ($M_{\rm b}=6.6\pm1.1$ $M_{\rm \oplus}$) and planet c ($M_{\rm c}=3.1^{+1.3}_{-1.2}$ $M_{\rm \oplus}$). The Doppler signal due to K2-3\,d remains undetected, likely because of its low amplitude compared to the radial velocity signal induced by the stellar activity. The closeness of the orbital period of K2-3\,d to the stellar rotation period could also make the detection of the planetary signal complicated. Based on our ability to recover injected signals in simulated data, we tentatively estimate the mass of K2-3\,d to be $M_{\rm d}$=2.7$_{\rm -0.8}^{\rm +1.2}$ $M_{\rm \oplus}$. These mass measurements imply that the bulk densities and therefore the interior structures of the three planets may be similar. In particular, the planets may either have small H/He envelopes (<1$\%$) or massive water layers (with a water content $\geq50\%$ of their total mass) on top of rocky cores. Placing further constraints on the bulk densities of K2-3\,c and d will be difficult; in particular, even by adopting a semester of intense, high-cadence radial velocity observations with HARPS-N and HARPS we would not have been able to detect the Doppler signal of K2-3\,d.}
   {}

   \maketitle
%

\section{Introduction}
\label{intro}
In the last decade, the search for potentially habitable exoplanets has focused particularly on M-dwarf stars. In this context, we regard a potentially habitable exoplanet as, in the broadest sense, one that orbits within or close to the habitable zone (HZ) of the parent star. Using the two main indirect detection methods - photometric transits and radial velocity (RV) - potentially habitable exoplanets can be detected more easily around M dwarfs than around earlier type stars. 

However, even M dwarfs present some challenges due to their faintness, their magnetic activity, and the difficulty in measuring accurate stellar parameters, which are essential if we wish to accurately determine the planetary parameters. Several exoplanet surveys have been devised to specifically target M dwarfs, such as the current ground-based photometric experiments MEarth \citep{irwin15}, APACHE \citep{sozzetti13} and TRAPPIST \citep{gillon17}, the upcoming SPECULOOS\footnote{http://www.speculoos.ulg.ac.be/cms/c$\_$3272698/en/speculoos-portail} and ExTrA \citep{bonfils15} projects, and surveys exploiting high resolution and high stability spectrographs (e.g. \citealt{bonfils13}; \citealt{delfosse13}; \citealt{affer16} and the HADES paper series; \citealt{quirrenbach16}). 

Results from the \textit{Kepler} and \textit{K2} missions have also been used to provide estimates for the occurrence of planets in the HZ of M dwarfs (e.g. \citealt{dressing13,dressing15,howell14,dressing17a,dressing17b}). These analyses suggest that small, low-mass planets are abundant around M dwarfs, and a significant fraction may have properties suitable for the emergence of life \citep{dressing15,tuomi14}. Space-based missions planned for the very near future, such as TESS \citep{tess14}, CHEOPS \citep{fortier14}, and PLATO \citep{rauer14} will target bright and nearby M dwarfs to detect potentially habitable planets, while JWST \citep{beichman14} and ground-based 25-40 meter class telescopes will characterize these planets' atmospheres. 

Some of the most intriguing exoplanet discoveries of the last few years are temperate 
planets around nearby M dwarfs, which orbit close to, or within, the predicted circumstellar HZ. For example, temperate, low-mass planets have been discovered around a number of nearby M-dwarf stars including Proxima Centauri \citep{anglada16}, TRAPPIST-1 \citep{gillon17}, LHS 1140 \citep{dittman17}, GJ 273 \citep{astudillo17}, K2-18 \citep{cloutier17}, GJ 667 C \citep{anglada13,delfosse13b,feroz14}, and Ross 128 \citep{bonfils17}. These new discoveries have stimulated many theoretical studies addressing the habitability of planets orbiting M dwarfs (see, e.g., \citealt{shields16} for a review on this topic). Open questions regarding the true habitability of these temperate planets include the influence of the spectral energy distribution and the activity of M dwarfs on planetary atmospheres, as well as the effect of tidal locking. Potentially habitable planets around M dwarfs are likely subject to physical conditions very different to those experienced on Earth, and their properties may also depend strongly on the M-dwarf spectral sub-type. 

Indeed, M-dwarfs that host temperate rocky planets exhibit a wide range of physical properties that might influence habitability. GJ 273, Ross 128, LHS 1140, Proxima Centauri and TRAPPIST-1 are mid-to-late M dwarfs (M3.5V, M4V, M4.5V, M5.5V and M8V, respectively). GJ 273 and Ross 128 have weak magnetic activity, while Proxima Centauri and TRAPPIST-1 have high activity levels, which will probably have a large impact on the potential habitability of their planets. K2-18 is, instead, an earlier-type M2.5 dwarf with low chromospheric activity, and GJ 667 C is a quiet M1.5 dwarf. The fact that these temperate planets are located at different distances from host stars spanning different spectral subtypes makes their comparative characterization particularly interesting. Unfortunately, many of these planets either do not transit, or their parent star is too faint to permit detailed follow-up, making it difficult to robustly characterize these planets and hence study their interior structures and compositions.          

Within this context, the planetary system around K2-3 (EPIC 201367065), a nearby (45 pc) M0 dwarf (V=12 mag; J=9.4 mag), presents an interesting opportunity for follow-up studies. Observations from the K2 mission revealed that K2-3 hosts at least three transiting small planets \citep{crossfield15}: K2-3\,b (R$_{\rm p}$=2\,R$_{\rm \oplus}$, P$_{\rm orb}$=10 days); K2-3\,c (R$_{\rm p}$=1.7\,R$_{\rm \oplus}$, P$_{\rm orb}$=24.6 days); K2-3\,d (R$_{\rm p}$=1.6\,R$_{\rm \oplus}$, P$_{\rm orb}$=45.5 days). According to the optimistic HZ boundaries derived by \cite{koppa13, koppa14}, planet K2-3\,d orbits close to the inner edge of the HZ of its host star. A particularly intriguing property of the K2-3 system is that the host star is bright enough to estimate the masses of these planets using existing high-resolution stabilized spectrographs. 

Measuring the masses of the K2-3 planets would be interesting for several reasons. First, by determining the mass and bulk density of the temperate planet K2-3d, we can extend the study of planets orbiting close or within the HZ to earlier-type host stars than those discussed earlier. Moreover, due to their measured sizes, planets K2-3\,c and K2-3\,d are optimal targets to test the results of \cite{rogers15}, who found that the majority of the observed planets with radius R$_{\rm p}\geq$1.6\,R$_{\rm \oplus}$ have densities too low to have a bulk rocky composition. \cite{fulton17} cast light on this result by finding evidence for a bimodal distribution of small planet sizes orbiting stars with $T_{\rm eff}$>4\,500 K with periods less than 100 days. Planets are preferentially found with radii around $\sim$1.3\,R$_{\rm \oplus}$ and $\sim$2.4\,R$_{\rm \oplus}$, with a region between 1.5 and 2.0 R$_{\rm \oplus}$ where planet occurrence is rare. This bi-modality likely reflects a separation between purely rocky planets and rocky cores surrounded by varying amounts of lower density volatile material. K2-3\,c and K2-3\,d fall within the ``gap'' between the two peaks in the planet radius distribution and orbit a star cooler than 4\,000 K, making them interesting test cases for understanding the composition of planets in this size regime at lower stellar irradiation than most of the planets analysed by \cite{fulton17}. 

Previously, other groups have recognized the appeal of the K2-3 system, and have begun conducting radial velocity follow-ups \citep{almenara15,dai16}. \cite{almenara15} found that stellar activity has a major impact on RV observations of K2-3 and, based on their single semester of monitoring, were unable to robustly measure the masses of K2-3\,c and K2-3\,d. Accurate determination of the masses of K2-3\,c and K2-3\,d evidently requires a larger dataset and denser sampling to trace out the activity signal. 

Here we present the results of an intense RV follow-up of K2-3 conducted over three seasons with the HARPS-N and HARPS spectrographs. Despite having data from several independent teams, and having precise transit ephemeris from high-precision photometry \citep{beichman16}, measuring the mass of K2-3\,c and K2-3\,d has been challenging.

The paper is organized as follows.
We first introduce the RV datasets and discuss the significant signals present in the data through a frequency analysis (Sect. \ref{sect:dataset}). In Sect. \ref{Section:star} we present updated stellar parameters and analyse the photometric light curves of K2-3 and the H$\alpha$ line activity indicator time series. The analysis of the RVs with Gaussian processes is described in Sect. \ref{Section:GPanalysis}. The significance of our best-fit solution for the masses of K2-3\,c and K2-3\,d is investigated through Monte Carlo simulations and is discussed in Sect. \ref{Section:simul}. Finally, we present the mass-radius diagram for the K2-3 planets and discuss the implications of our findings for their bulk composition.    


\section{Description and first look on the HARPS(-N) radial velocity datasets}
\label{sect:dataset}
Northern and Southern spectroscopic observations of K2-3 were carried out between January 22, 2015 and July 7, 2017, producing a total of 211 HARPS-N and 138 HARPS spectra.  Of the 349 spectra, 283 are unpublished observations while 66 HARPS observations were previously published by \cite{almenara15}. The HARPS-N spectra come from two independent programs: the HARPS-N Collaboration Guaranteed Time Observations (GTO)\footnote{https://plone.unige.ch/HARPS-N/science-with-harps-n} and the Global Architecture of Planetary Systems programs (GAPS, \citealt{benatti16}). The two collaborations shared observing time on this target to maximize the number of RV measurements and to optimise the observing strategy. The spectra were reduced with the version 3.7 of the HARPS-N Data Reduction Software (DRS) \citep{cosentino14}. 
\noindent The extraction and wavelength-calibration of the HARPS spectra were performed using the on-line pipeline \citep{lovis07}. The exposure time was fixed at 1800 s for both instruments, producing a typical SNR=30 at a wavelength $\lambda\sim$550 nm.

\subsection{Definition of the final dataset}
\label{sect:finaldataset}
We excluded some observations from our analysis for two reasons. We did not consider spectra with a signal-to-noise ratio S/N $\leq$11 as measured at $\lambda$=550 nm (echelle orders 46 and 49 for HARPS-N and HARPS, respectively). We then identified RV measurements potentially contaminated by scattered moonlight. This is particularly important for faint targets like K2-3 which lie close to the Ecliptic plane. By using the procedure described in Section 2.1 of \cite{malavolta17}, we identified 5 HARPS-N spectra that are probably contaminated by scattered moonlight, and we discarded them from the dataset\footnote{They correspond to the epochs BJD$_{\rm UTC}$ 2457407.638254, 2457412.711410, 2457412.732546, 2457413.703766, and 2457413.724936.}. We could not perform the same analysis on the full HARPS dataset because nearly half of those spectra were acquired with simultaneous Fabry-Perot and not with fibre B on sky. Nonetheless, we adopted a heuristic approach\footnote{We identified potentially contaminated observations as those where the absolute difference between the radial velocity of the Moon (which we assumed was equal to the Earth's barycentric radial velocity) and the target star was less than 15 km s$^{\rm -1}$ and more than 99$\%$ of the Moon's disk was illuminated.} to identify 4 potentially contaminated measurements, which we removed from the final dataset\footnote{We discarded observations on the epochs BJD$_{\rm UTC}$ 2457056.724361, 2457056.842204, 2457057.713776, and 2457057.849294.}. 

\noindent After these cuts, our final dataset includes 197 HARPS-N spectra and 132 HARPS spectra. 

\subsection{Radial velocity extraction}
\label{sect:rvextraction}
In this study we primarily use RVs extracted with the TERRA pipeline \citep{anglada12}. TERRA is commonly used to extract radial velocities of M-dwarfs because it typically provides measurements with better precision and lower scatter than those extracted for low-mass stars with the CCF recipe of on-line DRS pipeline \citep{perger17}. The HARPS fibre link was upgraded with octagonal fibres on May 28, 2015 \citep{locurto15}, introducing an RV offset between data acquired before and after the upgrade. We accounted for this offset by producing \textit{pre-} and \textit{post-upgrade} spectral templates to extract the RVs, and by introducing a velocity zero point offset for each dataset as free parameter when fitting the time series.

The TERRA RV time series are listed in the on-line Tables \ref{Table:radvel1} and \ref{Table:radvel2} for HARPS-N and HARPS, respectively. The median values of the internal errors are 1.88 \ms for HARPS-N and 2.04 \ms for HARPS.

\subsection{Preliminary RV frequency analysis}
\label{sect:rvgls}
\begin{figure}
\centering
\includegraphics[width=9cm]{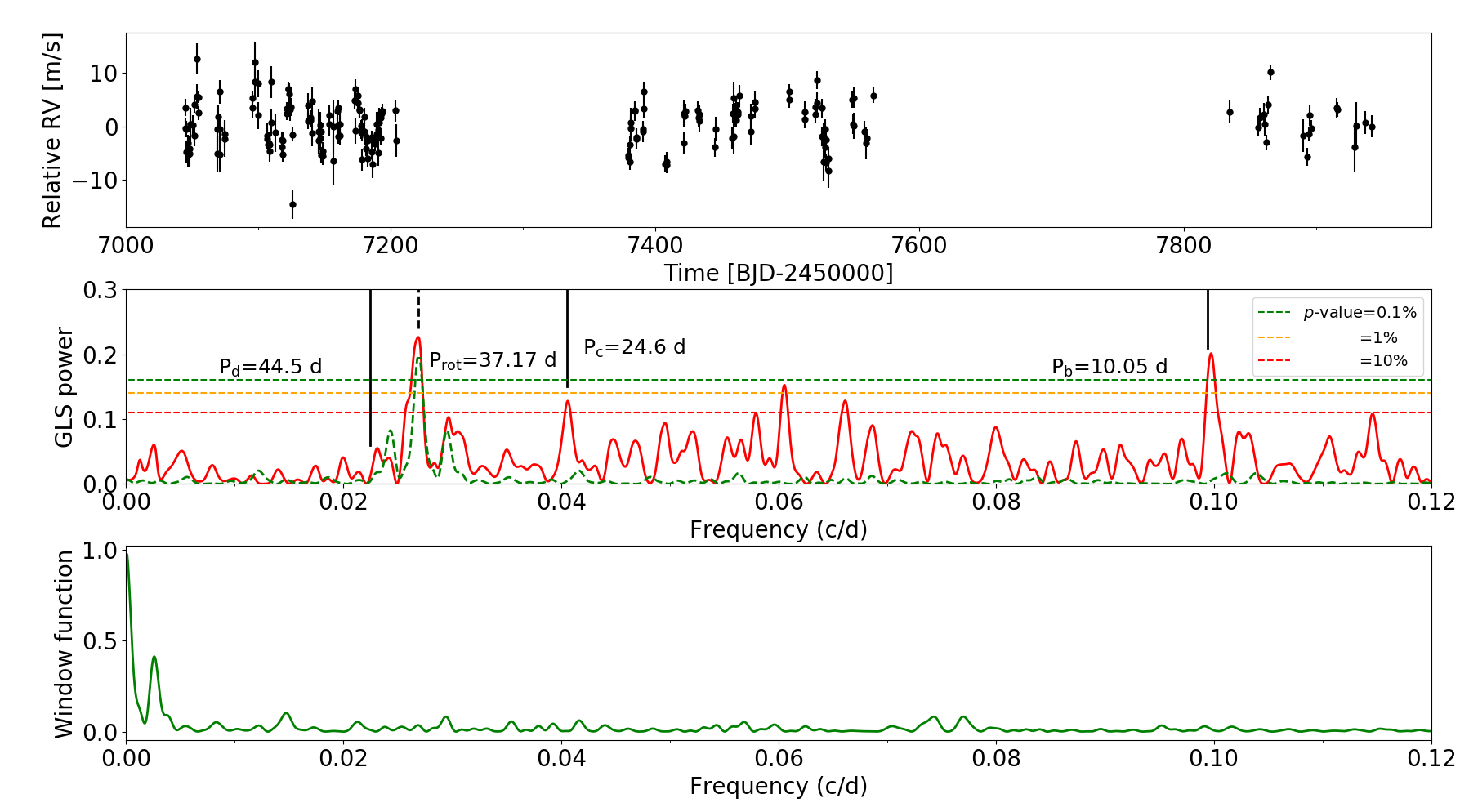}
\caption{\label{fig:rvHN} \textit{Upper plot}. Time series of the K2-3 radial velocities extracted with the pipeline TERRA using the HARPS-N spectra. \textit{Middle plot}. GLS periodogram of the RV time series (red line). Three levels of \textit{p}-values are indicated by the horizontal dashed lines. The green dashed line over plotted on the RV periodogram represents the window function of the measurements (shown in the \textit{bottom plot}) shifted in frequency to be superimposed on the strongest peak of the RV periodogram. This helps to identify the alias frequencies of the most relevant peaks. The stellar rotation and planetary orbital frequencies are indicated by vertical lines and corresponding labels.}
\end{figure}

\begin{figure}
\centering
\includegraphics[width=9.5cm]{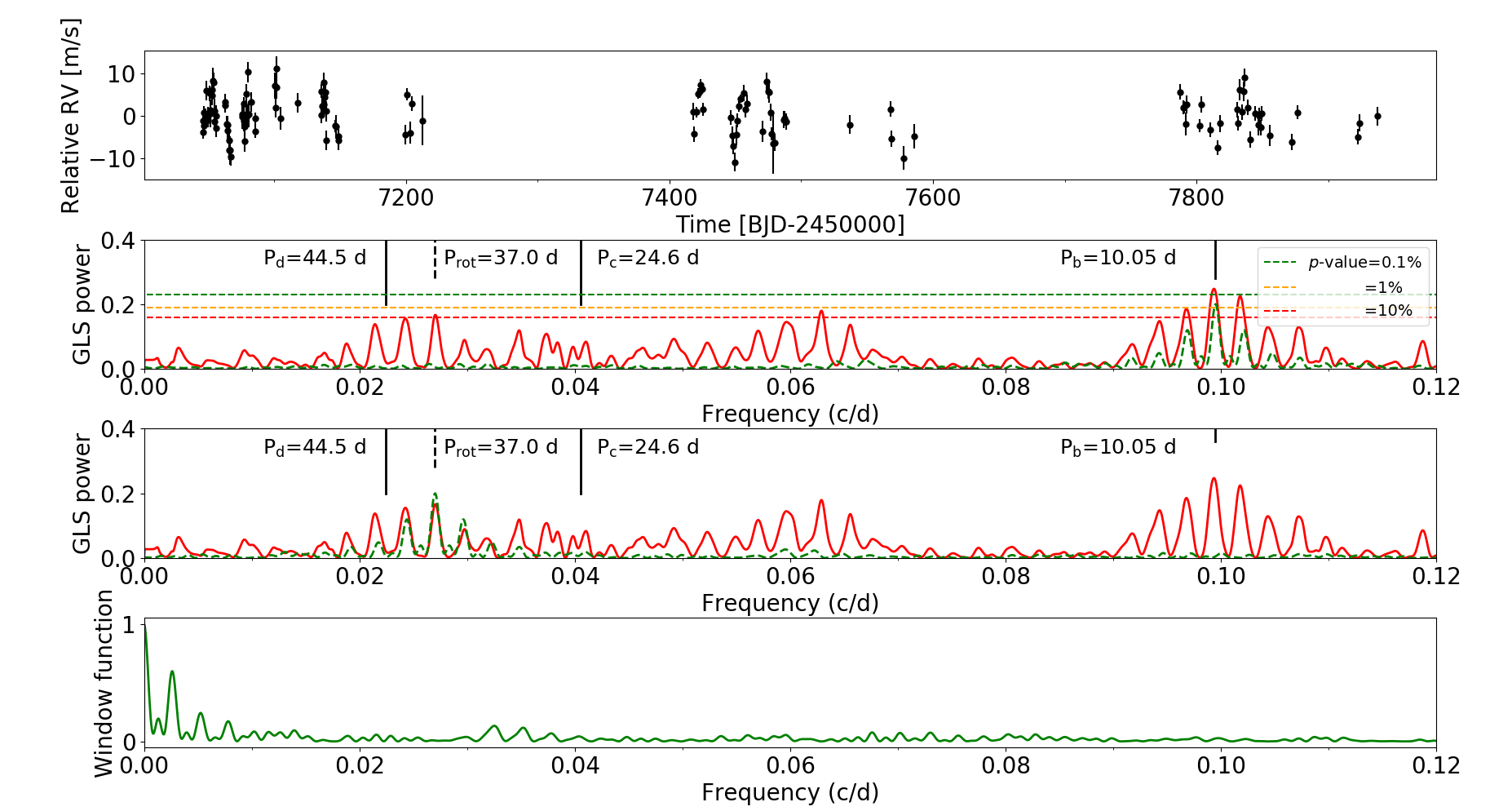}
\caption{ \label{fig:rvHS} As in Fig. \ref{fig:rvHN} but for TERRA RVs extracted from the HARPS spectra. \textit{Pre-} and \textit{post-}upgrade RV offsets have been applied, as derived from our analysis described in Sect. \ref{Section:GPanalysis}.}
\end{figure}

\begin{figure}
\centering
\includegraphics[width=9.5cm]{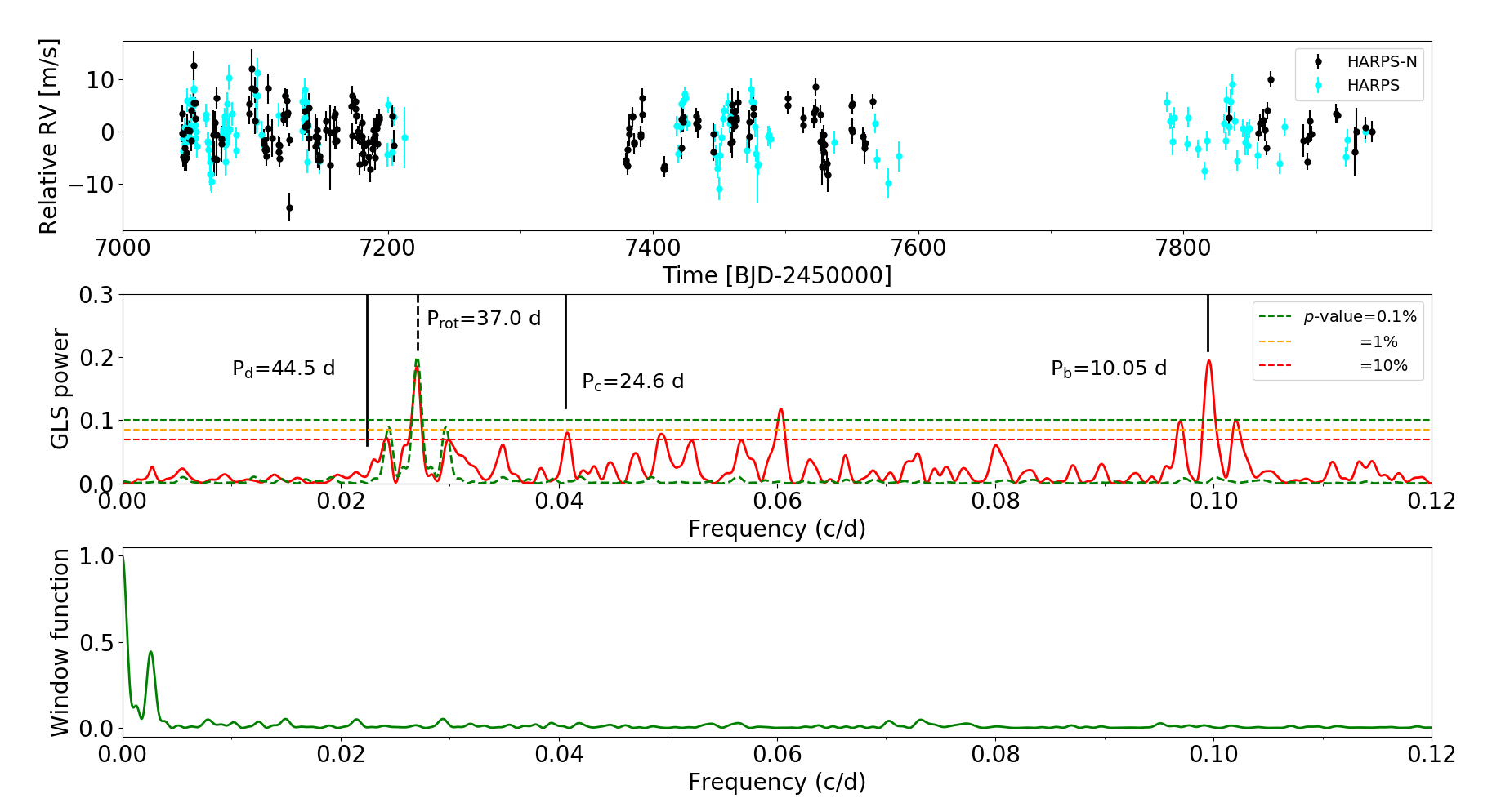}
\caption{\label{fig:rvHNHS} As in Fig. \ref{fig:rvHN} but for all TERRA HARPS-N and HARPS RVs. We have applied an offset to each separate dataset, as derived from our analysis described in Sect. \ref{Section:GPanalysis}.}
\end{figure}

We conducted a preliminary analysis of the radial velocity datasets using the Generalized Lomb-Scargle (GLS) algorithm \citep{zech09} to identify significant signals. We performed the GLS frequency analysis of the Northern and Southern datasets separately as well as the combined RV dataset, removing RV offsets as determined by the analysis described in Sect. \ref{Section:GPanalysis}. We show the resulting periodograms in Figs. \ref{fig:rvHN}-\ref{fig:rvHNHS}. 

The HARPS-N periodogram is dominated by a signal at 37.2$\pm$0.1 days\footnote{In this work we adopt the uncertainties calculated with GLS as formal errors of the periods.}, which we identify as the rotation period of the star (see Sect. \ref{Section:star}). This signal has a \textit{p}-value=0.01$\%$, as estimated by bootstrapping the data, i.e. by randomly drawing the RV measurements (with replacement) and generating 10\,000 mock datasets. Its semi-amplitude is 2.9$\pm$0.3 \ms, as estimated with GLS, slightly lower than the RMS of the data (4.1 \ms). A significant peak with a slightly lower power appears at the orbital period of K2-3\,b, while the peak corresponding to the orbital period of K2-3\,c appears with less significance (\textit{p}-value$\sim$1$\%$). The orbital period of K2-3\,d is undetected in the HARPS-N periodogram.

The HARPS periodogram is dominated by the orbital period of K2-3\,b (Fig. \ref{fig:rvHS}; \textit{p}-value=$0.01\%$). The signal produced by the stellar rotation has a \textit{p}-value around $1\%$, therefore it is much less significant than in the HARPS-N data. The window function is responsible for a pattern of alias frequencies around these signals. In addition to the one-\textit{year} aliases, one-\textit{month} aliases (synodic month frequency f$_{\rm s.m.}$= 0.03386 c/d) are also probably present, which is expected for a star near the ecliptic. For example, a one-\textit{month} alias of the 37-day signal occurs at 0.061 c/d (=1/37+f$_{\rm s.m.}$ c/d), while a one-\textit{month} alias of P$_{\rm b}$ occurs at 0.065 c/d. Both aliases are visible in Fig. \ref{fig:rvHS}. 

The signals that most clearly emerge in the periodogram of the combined dataset are those of the planet K2-3\,b and the stellar rotation period, which are highly significant and have almost the same power (Fig. \ref{fig:rvHNHS}). The signal of planet K2-3\,c is more weakly present, with a \textit{p}-value$\sim$1$\%$.

\section{Stellar parameters and activity}
\label{Section:star}

The main properties of the star K2-3 are summarized in Table \ref{Tab:starparam}. While K2-3 has previously been characterized by \cite{crossfield15} and \cite{almenara15}, we independently determined the stellar parameters from our HARPS-N spectra using the method developed by \cite{maldo15}\footnote{https://github.com/jesusmaldonadoprado/mdslines}. We estimated the star's effective temperature and iron abundance using measurements of the pseudo-equivalent widths of the spectral features. We then determined stellar mass, radius, and surface gravity using empirical relationships. Our results are consistent with, but slightly more precise than, previous estimates. We adopt these new values for the analysis in the rest of the paper.

\begin{table}
  \caption{Stellar parameters for K2-3}
         \label{Tab:starparam}
         \centering
   \begin{tabular}{l l l}
            \hline
            \noalign{\smallskip}
            Parameter  &  Value & Ref. \\
            \noalign{\smallskip}
            \hline
            \noalign{\smallskip}
            RA [deg, ICRS J2015] & 172.3353579 &  \\
            DEC [deg, ICRS 2015] & -1.4551256 &  \\
            Mass [$\msun$] & 0.62$\pm$0.06 & (1) \\
            & 0.60$\pm$0.09 & (2) \\
            & 0.61$\pm$0.09 & (3) \\	
            & 0.60$\pm$0.09 & (4) \\
            Radius [$\rsun$] & 0.60$\pm$0.06 & (1) \\
            & 0.56$\pm$0.07 & (2) \\
            & 0.55$\pm$0.04 & (3) \\
            & 0.56$\pm$0.07 & (4) \\
            Effective temperature, T$_{\rm eff}$ [K] & 3835$\pm$70 & (1) \\
            & 3896$\pm$189 & (2) \\
            $\feh$ & -0.01$\pm$0.09 & (1) \\
            & -0.32$\pm$0.13 & (2) \\
            Surface gravity, $\logg$ [log$_{\rm 10}$(cgs)] & 4.66$\pm$0.05 & (1) \\
            & 4.73$\pm$0.06 & (3) \\
            & 4.72$\pm$0.13 & (4) \\ 
            Density [$\rhosun$] & 3.51$\pm$0.61 & (3) \\ 
            $\log(L/\lsun$) & -1.15$\pm$0.09 & (1) \\
            Age [Gyr] & $\geqslant$1 & (2) \\
            \noalign{\smallskip}
            \hline
     \end{tabular}  
     \tablefoot{(1) This work: derived from HARPS-N spectra using the method described by \cite{maldo15}.
     (2) \cite{crossfield15}. (3) \cite{almenara15}. (4) \cite{sinukoff16}.}
\end{table}

\subsection{Photometry}
\label{sec:photometry}
\begin{figure}
\centering
\includegraphics[width=9cm]{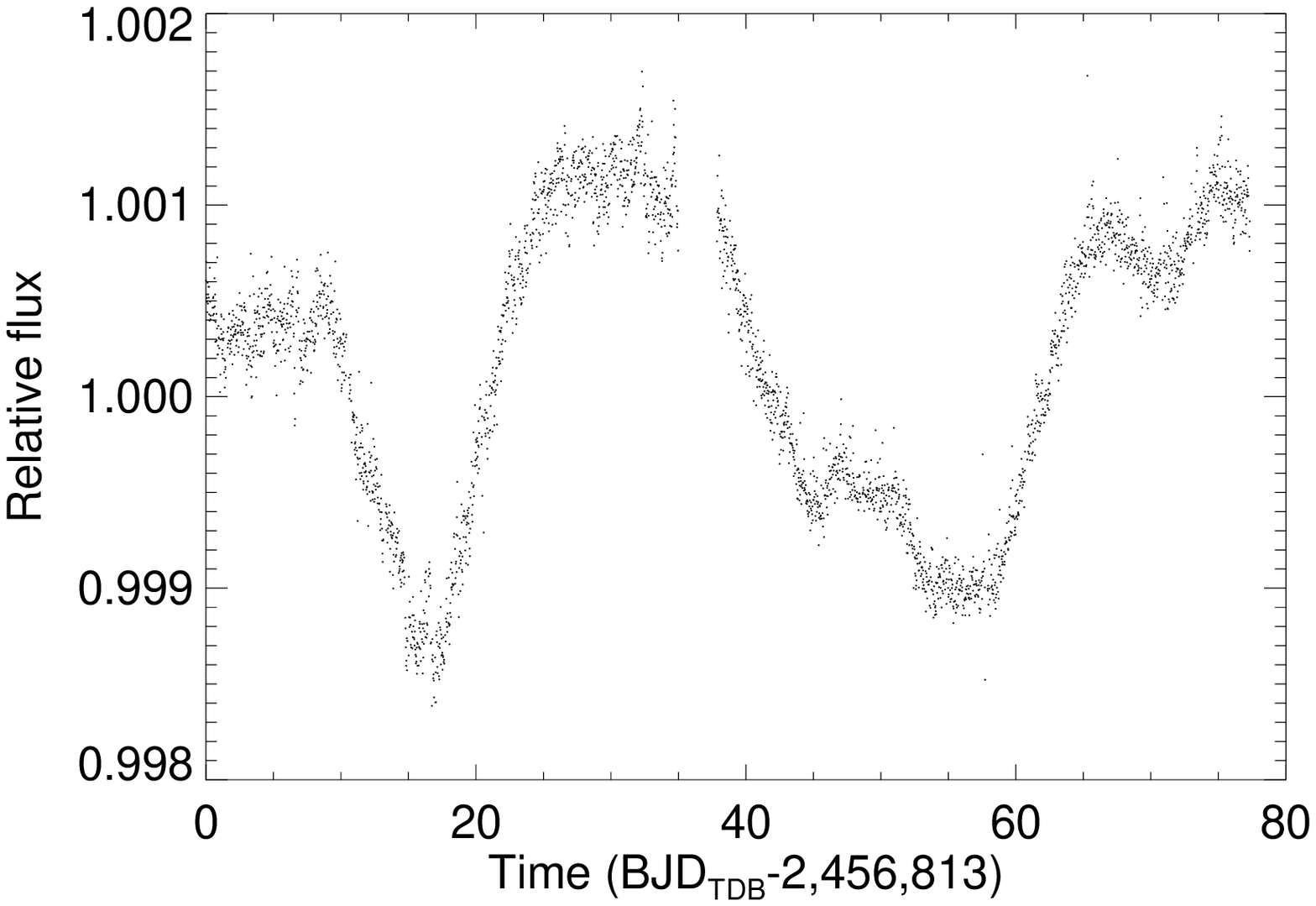}
\includegraphics[width=9cm]{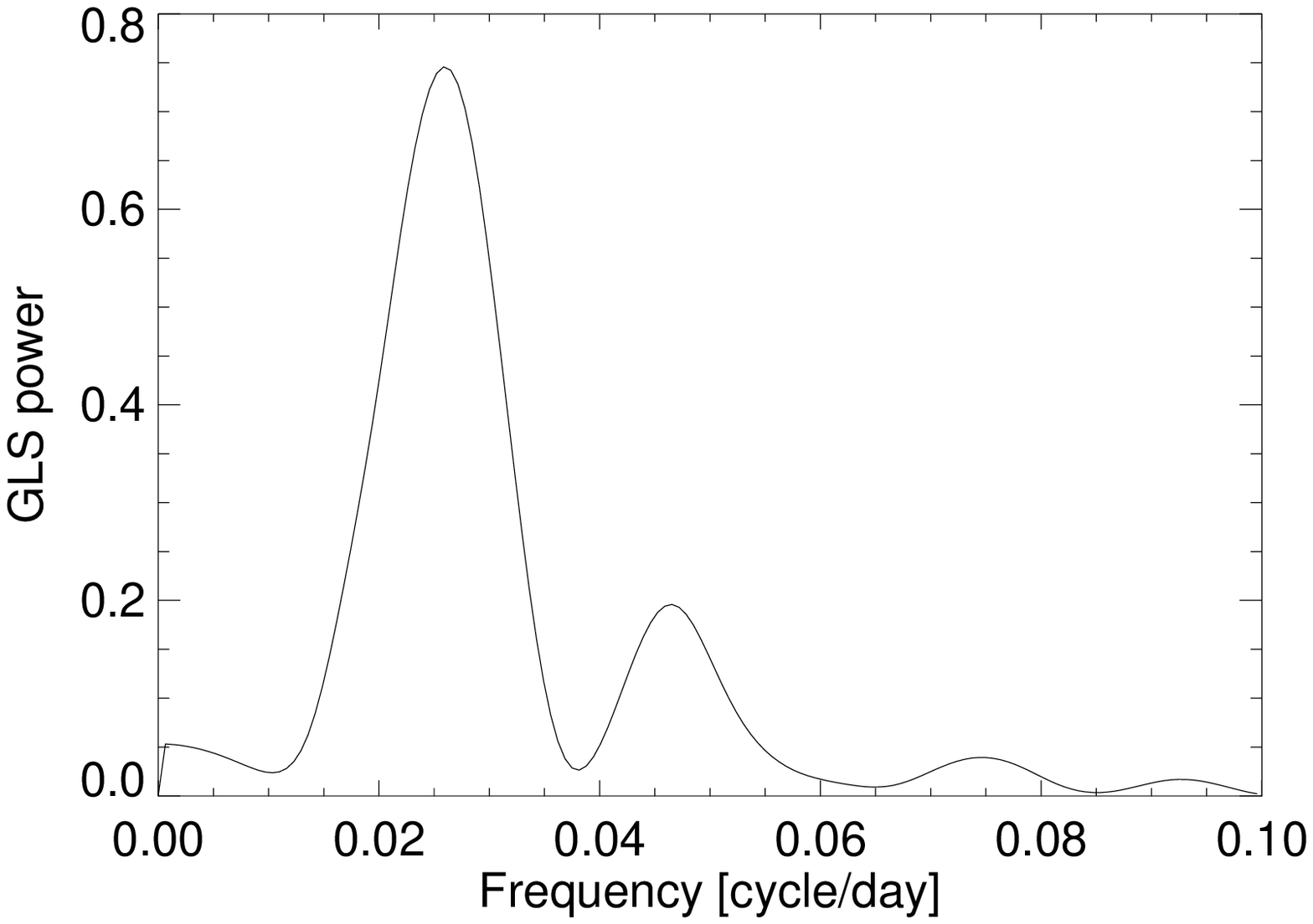}
\caption{\label{fig:k2lightcurve} (\textit{Top}) \textit{K2} light curve of K2-3, with the planetary transits removed. (\textit{Bottom}) GLS periodogram of the binned light curve (one point per day), showing a peak at $P_{\rm rot}$=38.3 days.} 
\end{figure}

The \textit{K2} light curve of K2-3, with all the transit signals removed, is shown in Fig. \ref{fig:k2lightcurve}. We have used the K2SFF\footnote{https://archive.stsci.edu/prepds/k2sff/} light curve processed as described by \cite{vanderjohns14} and \cite{vander16}. The light curve shows a quasi-periodic modulation with a flux semi-amplitude of $\sim0.1\%$. There is also clear evidence for changes from one rotation to the next, likely due to the evolution of active regions. In the bottom panel of Fig. \ref{fig:k2lightcurve} we show the GLS periodogram of the binned K2 light curve (one point per day), which shows a strong peak at $P_{\rm rot}$=38.3$\pm$0.7 days\footnote{ Since the K2 data barely cover two rotation cycles of K2-3, this estimate should not be considered particularly accurate.}. Even though the \textit{K2} light curve was obtained about six months before the first spectroscopic observations and only covers about two rotation periods, the high signal-to-noise photometry is useful for constraining the stellar rotation period. It is interesting to note that, while the amplitude of the star's photometric rotational variability is fairly low, the stellar rotation frequency is nonetheless the strongest signal in the star's radial velocity time series. 

\begin{figure}
\centering
\includegraphics[width=9cm]{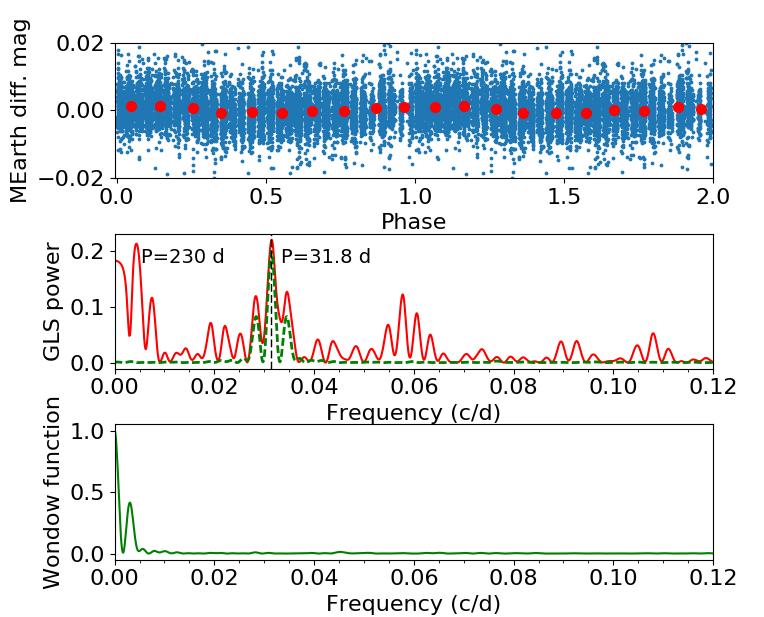}
\caption{\label{fig:MElightcurve} (\textit{Top}). Light curve of K2-3 from the MEarth-south survey, folded at the best period P=31.8 days found using GLS. (\textit{Middle}). GLS periodogram of the MEarth light curve. The green line corresponds to the window function of the measurements (\textit{Bottom plot}), which is shifted in frequency so that the peak superimposes on the major peak of the RV periodogram, in order to directly identify alias frequencies.}
\end{figure}

We also obtained time-series photometry of K2-3 from the MEarth survey \citep{irwin15}. We monitored K2-3 with one of the 40-cm telescopes of the Southern MEarth array from January 21, 2015 to April 4, 2016. With a total of 8669 data points, the light curve overlaps with the first two seasons of the spectroscopic observations. The fact that the MEarth data are contemporaneous with many of our radial velocity observations and their high photometric precision (due in part to the excellent observing conditions at Cerro Tololo), make the MEarth light curve potentially useful for characterizing the stellar activity. We analysed the MEarth data (in nightly bins) by calculating a GLS periodogram. The highest peak in the periodogram is at 31.8$\pm$0.2 days, with semi-amplitude of $\sim$1 mmag (Fig. \ref{fig:MElightcurve}). This period does not correspond to the more robust stellar rotation period derived from the uninterrupted observations of \textit{K2} or using the  H$_{\rm \alpha}$ spectroscopic indicator (Sect. \ref{Section:actind}); nonetheless the result is worth mentioning because the data have been collected with uneven sampling by a ground-based small-aperture telescope, and they show very low amplitude modulation compared to the typical photometric error. 
The periodogram also shows an additional peak at a period of about $\sim$230 days, without any counterpart in the window function. While the signal could be due to systematics present in the MEarth data, its nature is unclear, and an astrophysical origin cannot be ruled out (see Sect. \ref{Section:actind}). 

\subsection{Spectroscopic activity indexes}
\label{Section:actind}
\begin{figure}
\centering
\includegraphics[width=0.5\textwidth]{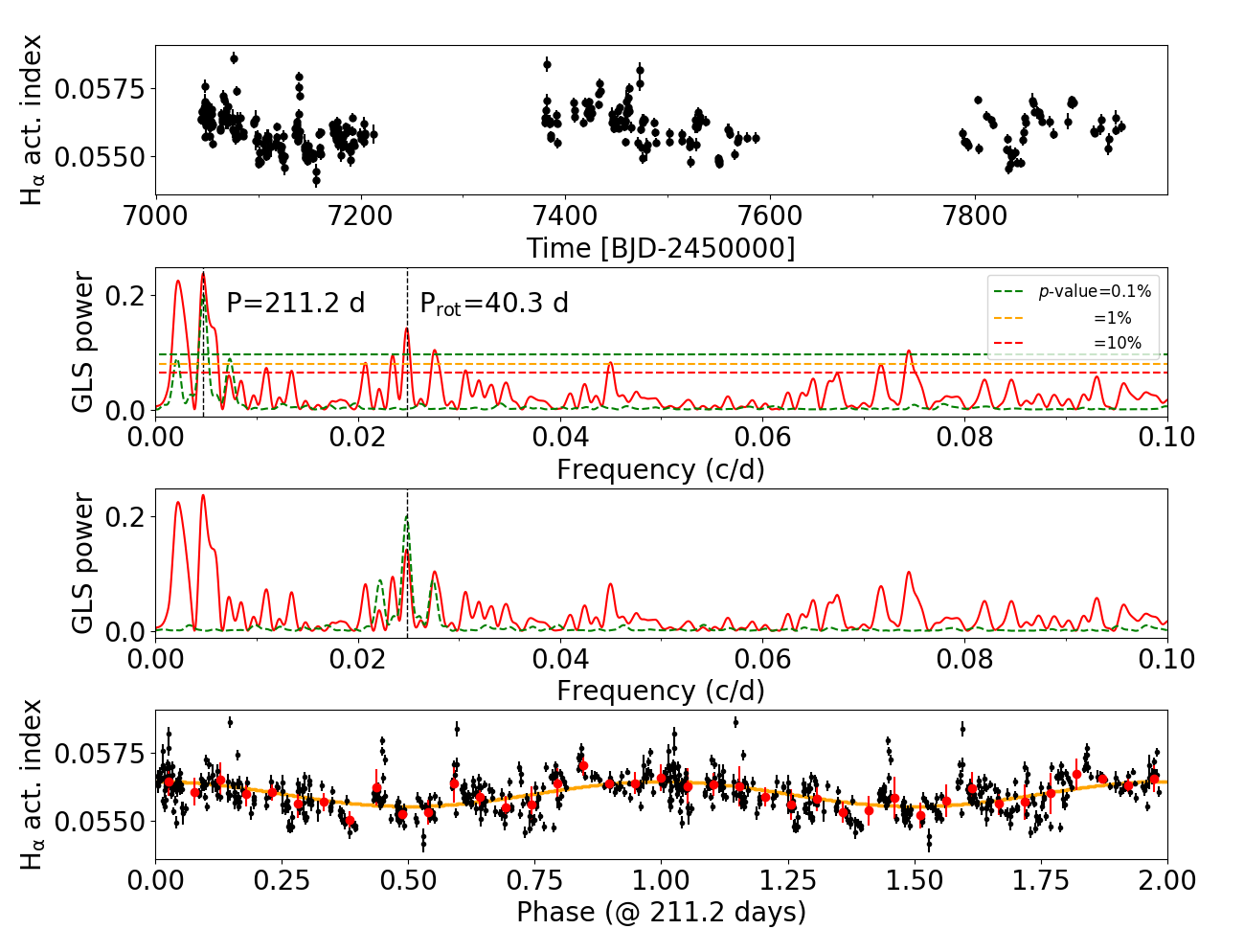}
\caption{\label{fig:actindic} (\textit{Top}) Time series of the activity indicator based on the H$\alpha$ line extracted from the HARPS and HARPS-N spectra. (\textit{Second and third plot}) GLS periodogram of the dataset. The dashed horizontal lines mark the p-value levels as derived from a bootstrap analysis. The green curve corresponds to the window function of the measurements, which is shifted in frequency so that the peak superimposes on the major peak of the RV periodogram (second plot), to directly identify alias frequencies. In the third plot the window function is shifted to superimpose on the P$\sim$40 days rotational signal. (\textit{Bottom}) Time series phase-folded at the period P=211 days. Red points represent the average of the data within 20 bins in the phase range [0,1].}
\end{figure}

We studied the activity level of K2-3 during the time period of our observations using the H$\alpha$ line as a spectroscopic activity indicator, which we extracted from our spectra using the method described by \cite{gomesdasilva11}\footnote{We did not analyse the activity indicator based on the CaII H$\&$K lines because of the very low S/N in this spectral region.}.  The time series of the H$\alpha$ activity indicator is shown in Fig. \ref{fig:actindic}, and the data are listed in Table \ref{Table:halpha}. We analysed the time series by calculating a GLS periodogram to identify periodicities related to a possible activity cycle and stellar rotation. The GLS periodogram is shown in the second and third panels of Fig. \ref{fig:actindic}.

The highest peak in the periodogram occurs at P=211$\pm$3 days, with a power comparable to that of its 1-year alias frequency at $\sim450$ days. The existence of a long-term modulation is especially clear after looking at the data of the first two seasons. A bootstrap (with replacement) Monte Carlo analysis based on 10\,000 mock datasets reveals that these peaks are statistically significant, with false alarm probabilities lower than 0.1$\%$. We are, however, unable to ascertain whether the 211 day period or its alias at 450 days is the true underlying period. The origin of this long-period signal is unclear. If the signal is astrophysical, one possible explanation is an intermediate-duration activity cycle. Some tentative evidence exists for such cycles in low-mass stars \citep{Savanov2012,robertson13}, and they could represent sub-cycles superimposed on longer duration activity cycles, as observed for the Sun (so called `Rieger cycles`). We note that the 211 day period is close to the 230 day signal observed in the MEarth photometry.

\begin{figure}
\centering
\includegraphics[width=0.5\textwidth]{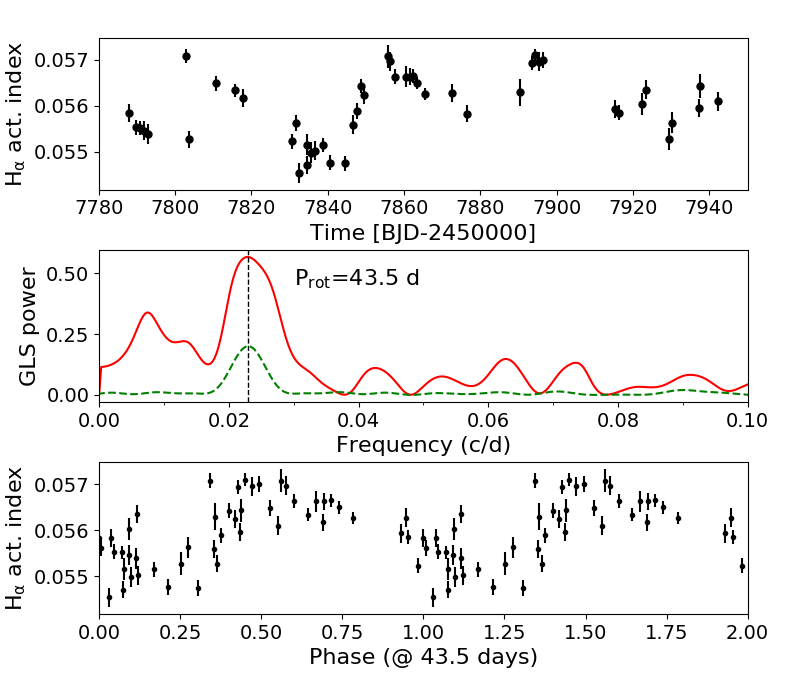}
\caption{\label{fig:actindic2} (\textit{Top}) Time series of the activity indicator based on the H$\alpha$ line extracted from the HARPS and HARPS-N spectra. Here, only the dataset of the third season is shown. (\textit{Middle}) GLS periodogram of the dataset. The green line corresponds to the window function of the measurements, which is shifted in frequency so that the peak superimposes to the major peak of the RV periodogram, to directly identify possible alias frequencies. (\textit{Bottom}). Time series phase-folded at the period P=43.5 days.}
\end{figure}

The second highest peak in the periodogram of the H$\alpha$ indicator is close to the expected stellar rotation period and is highly significant (P=40.3$\pm$0.1 days, \textit{p}-value<0.1$\%$). This signal is particularly strong in the last season of observations; during this time, a clear modulation related to $P_{\rm rot}$ is visible in the H$\alpha$ time series, which covers nearly four stellar rotations. A GLS analysis of only the last season of observations identifies a periodicity of 43.5$\pm$1.0 days. Folding the data at this period reveals that the modulation does not have a simple sinusoidal shape (Fig. \ref{fig:actindic2}). 

\section{Gaussian Process regression analysis of radial velocities}
\label{Section:GPanalysis}
We used the stellar activity information derived in Section \ref{Section:star} to perform a detailed analysis of the combined HARPS-N and HARPS RV datasets within a Bayesian framework based on Gaussian process (GP) regression. 
It has now become standard in RV analysis to use GPs to model the stellar contribution to the RV variations, jointly with a number of Keplerian functions describing the planetary orbital motion (see \citealt{haywood14} for the first application of this technique). This approach has proven to be a powerful way of retrieving planetary masses from high-precision RV measurements when a signal closely related to the stellar rotation period, or its harmonics, is present in the RV time series \citep{dumusque17}.

For a general description of the GP method, and its performances when applied to RV time series, we refer, amongst others, to the recent works by \cite{morales16}, \cite{cloutier17}, \cite{damasso17}, and \cite{dittman17}. The GP regression is based on the choice of a specific kernel; i.e. a covariance matrix describing the correlation between measurements taken at two different epochs. 
For the case of K2-3, the so-called \textit{quasi-periodic} (q-p) kernel is particularly useful because a signal very likely related to the stellar rotation period P$_{\rm rot}$ dominates the RV time series (Section \ref{sect:dataset}), and there is evidence for an evolutionary time scale of the active regions close to P$_{\rm rot}$ (Sect. \ref{sec:photometry} and \ref{Section:actind}). 
Each element of the covariance matrix has the form
{
\begin{eqnarray} \label{eq:1}
K(t, t^{\prime}) = h^2\cdot\exp\bigg[-\frac{(t-t^{\prime})^2}{2\lambda^2} - \frac{sin^{2}(\dfrac{\pi(t-t^{\prime})}{\theta})}{2w^2}\bigg] + \nonumber \\
+\, \bigg[\sigma^{2}_{\rm RV, instr}(t)\,+\,\sigma_{\rm jit, instr}^{2}\bigg]\cdot\delta_{t,t^{\prime}},
\end{eqnarray}
where $t$ and $t^{\prime}$ represent two different epochs. 
Here, the first term represents the quasi-periodic kernel, which is composed of a periodic term and an exponential decay term. This functional form is suitable for modelling a recurrent signal linked to stellar rotation, and takes into account the finite-lifetime of the active regions.
Eq. \ref{eq:1} contains four covariance matrix hyperparameters: $h$ represents the amplitude of the correlations; $\theta$ represents the rotation period of the star; $w$ is the length scale of the periodic component, linked to the size evolution of the active regions; and $\lambda$ is the correlation decay time scale, that can be physically related to the active regions lifetime.
The remaining parameters in Eq. \ref{eq:1} are: $\sigma_{\rm RV, instr}(t)$, which is the RV internal error at time \textit{t} for each spectrograph (or independent dataset); $\sigma_{\rm jit, instr}$, which are additional uncorrelated `jitter' terms, one for each instrument (or independent dataset), that we add in quadrature to the internal errors to account for additional instrumental effects and noise sources neither included in $\sigma_{\rm RV, instr}(t)$ nor modelled by the q-p kernel; and $\delta_{t,t^{\prime}}$, which is the Kronecker delta function.

Our GP analysis is based on a Markov chain Monte Carlo (MCMC) algorithm. The model, algorithms, and statistical framework used in this work are the same as described in \cite{damasso17}, and we refer the reader to that work for a detailed description. Here, we assumed circular orbits for all the planets. The best fit values and uncertainties for each jump parameter are calculated as the median of the marginal posterior distributions and the 16$\%$ and 84$\%$ quantiles. 

\subsection{Choice of the priors}
The priors adopted in our analysis are listed in Table \ref{Table:priorsmultidataset}. In this subsection, we describe and justify some of our choices for the priors. 

In our fits, we allow the semi-amplitude \textit{h} of the correlated stellar signal to vary up to a value of 5 \ms, which represents nearly twice the value estimated by GLS at P=37 days. 

The prior range used for the stellar rotation hyperparameter $\theta$ is defined on our rotation period estimates from the \textit{K2} photometry and the H$\alpha$ activity indicator. This also takes into account the results of a trial MCMC analysis that was run after the conclusion of the second observing season, which adopted a larger prior range and indicated that the posterior distribution was well constrained within the range [35,43] days.

For the active regions evolutionary time scale $\lambda$ we adopt a uniform prior between 20 and 60 days, corresponding nearly to $P_{\rm rot}$/2 and 1.5$\cdot$P$_{\rm rot}$. This choice is first motivated by the marginal posterior obtained with the trial MCMC analysis of the dataset for the first and second seasons, which is symmetric around $\lambda$=36 days ($\sigma\sim8$ days). An evolutionary time scale of the order of the stellar rotation period can also be guessed directly by looking at the \textit{K2} light curve, which shows changes in its pattern from one rotation to the next \footnote{Due to the short time baseline, we could not constrain $\lambda$ through the analysis of the autocorrelation function, as done by \cite{morales16} using the longer baseline of the Kepler photometry}. We also used the results of \cite{giles17} (Eq. 8) to get a loose estimate of the time scale based on the RMS of the \textit{K2} data and the stellar effective temperature. The equation for determining the time scale was derived by \cite{giles17} using a calibration sample not biased by spectral type and, though faster rotators than K2-3 are analysed in Giles et al. work, it could be tentatively used for a star with P$_{\rm rot}$=40 days. With an RMS = 7.6$\cdot10^{-4}$ mag and $T_{\rm eff}$=3835 K, we get a time scale of 38$^{+19}_{-13}$ days. 

For the planetary orbital parameters, we fixed the upper limit of the Keplerian semi-amplitude to 5 \ms for K2-3\,b and K2-3\,c, and to 3 \ms for planet K2-3\,d. For K2-3\,b, $K$=5 \ms is more than twice the value estimated by GLS after removing the stellar rotation signal, and represents a conservative upper limit also for the planet K2-3\,c. For the outermost planet, K2-3\,d, the absence of a signal in the GLS periodogram suggests that $K_{\rm d}$ should be significantly lower. The priors on the orbital period and time of transit were taken from \cite{beichman16}, who determined the ephemerides by combining K2 observations with additional transits observed with the \textit{Spitzer} space telescope.

\subsection{Analysis of the combined radial velocity dataset}
\label{Section:gpn+s}

We analysed the full RV dataset without binning the data when more than one observation is available during the same night. To guarantee a wide exploration of the parameter space, we adopted 150 independent chains properly initialized to start from well separated locations. We discarded the first 3\,000 steps of each chain by resetting the sampler. The MCMC chains reached the convergence according to the Gelman-Rubin statistics after 18\,000 steps, and a further burn-in (0.75$\%$ of the total steps) was applied to calculate the best-fit values of the parameters (see \citealt{eastman13} and references therein). 

We show the best-fit results for all the free parameters and derived quantities in Table \ref{Table:percentilesmultidata}.
The mass of the super-Earth K2-3\,b is robustly determined with a signal-to-noise ratio of $\sim$7$\sigma$, while the mass of K2-3\,c is detected with a lower significance of 2.6$\sigma$. The mass of K2-3\,d remains undetermined, with a 1$\sigma$ upper limit of about $\sim2\mearth$. This suggestes that the mass of K2-3\,d is too low to be detected, but it is also plausible that the low-amplitude of the Doppler signal, compared to that of the stellar activity component, and its proximity to the stellar rotation period allow the signal to be absorbed into our GP activity model \citep{vander16}. We investigate the robustness of this upper limit in Sections \ref{Section:simul1} and \ref{Section:simul2}. 

We show the RV curves in Fig. \ref{fig:foldedrv_terra_qp}, with the stellar activity contribution subtracted, and folded at the best-fit orbital periods. The stellar contribution to the RV time series resulting from our fit is shown in Fig. \ref{fig:stellarnoise_terra_qp}. 
The stellar activity component of the model is reliably described by our best-fit results. The stellar rotation period and the evolutionary time scale of the active regions appear to be well characterized for both datasets. 
As we have found from the analysis of the RV datasets of other targets, the GP quasi-periodic regression tends to suppress low frequencies in the residuals. This means that if an additional long-term modulation is actually present in the original data, and it is not explicitly included in the fitted model, it would be absorbed by the stellar activity component and disappear in the residuals, after removing the planetary solutions. However, we note that the data in Fig. \ref{fig:stellarnoise_terra_qp} do not show any long-term trend residual suggestive of an activity cycle or a longer period companion to K2-3 not included in our global model.   
We also note that the stellar rotation period, as retrieved by the GP regression, differs from that derived with GLS by $\sim$3 days. This difference is because the functional form of the quasi-periodic kernel used to model the stellar contribution is different from the simple sinusoid used by the GLS periodogram. We calculated the GLS periodogram of the GP quasi-periodic stellar activity timeseries and found the strongest peak at P=37 days, reproducing the period found in our GLS analysis of the RV dataset.  
The RMS of the residuals is 2.6 \ms, only slightly higher than the median of the internal errors. 

\subsection{Cross-check with the alternative RV extraction method}
\label{Section:naira}
In order to test the robustness of the results using RVs obtained with the TERRA dataset,
we also extracted the RVs using an independent pipeline described by \cite{astudillo15,astudillo17}.
In brief, this pipeline works by aligning all the spectra to a common reference frame by removing the Earth's barycentric radial velocity and the RVs of K2-3 measured by the DRS as an initial guess. Then a median template is computed from the aligned spectra and regions of the spectra contaminated by tellurics are rejected. The template is used to calculate a chi-square profile as a function of RVs for each individual spectrum, whose minimum corresponds to the stellar RV. 

The radial velocities from this alternative pipeline are listed in Tab. \ref{Table:radvel3}-\ref{Table:radvel4}. We performed the same analysis as described in Section \ref{Section:gpn+s} with the data from this alternative pipeline, and found that the results of this analysis (shown in Table \ref{Table:percentilesmultidata}) are in very good agreement with those obtained with TERRA. In particular, the planetary parameters are all in agreement within 1$\sigma$, strengthening our findings.


\begin{table}
  \caption[]{Prior probability distributions for the three-planet circular model parameters.}
         \label{Table:priorsmultidataset}
         \centering
   \begin{tabular}{l l}
            \hline
            \noalign{\smallskip}
            Jump parameter     &  Prior \\
            \noalign{\smallskip}
            \hline
            \noalign{\smallskip}
            $h$ [m$\,s^{-1}$] &$\mathcal{U}$ (0.5, 5)\\
            $\lambda$ [days] & $\mathcal{U}$ (20, 60) \\
            $w$ & $\mathcal{U}$ (0.001, 1) \\
            $\theta$ [days] & $\mathcal{U}$ (35, 43) \\
            $K_{\rm b}$ [m$\,s^{-1}$] & $\mathcal{U}$ (0.05, 5) \\
            $P_{\rm b}$  [days] & $\mathcal{N} (10.054544, 0.000029^{2})$ \tablefootmark{a}\\
            $T_{\rm 0,b}$  [BJD-2\,400\,000] & $\mathcal{N} (56813.42024, 0.00094^{2})$ \tablefootmark{a}\\
            $K_{\rm c}$ [m$\,s^{-1}$] & $\mathcal{U}$ (0.05, 5) \\
            $P_{\rm c}$  [days] & $\mathcal{N} (24.64638, 0.00018^{2})$ \tablefootmark{a}\\
            $T_{\rm 0,c}$  [JD-2\,400\,000] & $\mathcal{N} (56812.2777, 0.0026^{2})$ \tablefootmark{a}\\
            $K_{\rm d}$ [m$\,s^{-1}$] & $\mathcal{U}$ (0.05, 3) \\
            $P_{\rm d}$  [days] & $\mathcal{N} (44.55765, 0.00043^{2})$\tablefootmark{a} \\
            $T_{\rm 0,d}$  [JD-2\,400\,000] & $\mathcal{N} (56826.2248, 0.0038^{2})$ \tablefootmark{a} \\
            $\gamma_{\rm HARPS-N}$ [m$\,s^{-1}$] & $\mathcal{U}$ (-10, +10) \\
            $\gamma_{\rm HARPS-pre}$ [m$\,s^{-1}$] & $\mathcal{U}$ (-10, +10) \\
            $\gamma_{\rm HARPS-post}$ [m$\,s^{-1}$] & $\mathcal{U}$ (-10, +10) \\
            $\sigma_{\rm jit, HARPS-N}$ [m$\,s^{-1}$] & $\mathcal{U}$ (0.05, 5) \\
            $\sigma_{\rm jit, HARPS-pre}$ [m$\,s^{-1}$] & $\mathcal{U}$ (0.05, 5) \\
            $\sigma_{\rm jit, HARPS-post}$ [m$\,s^{-1}$] & $\mathcal{U}$ (0.05, 5) \\
            \noalign{\smallskip}
            \hline
     \end{tabular}  
     \tablefoot{
     \tablefoottext{a}{Ephemeris from Beichman et al. (2016), derived from transit observations with \textit{K2} and \textit{Spitzer}. $T_{\rm 0}$ is the time of inferior conjunction. } }
\end{table}

\begin{table*}
  \caption[]{Best-fit solutions for the quasi-periodic GP model applied to the combined HARPS/HARPS-N RV time series extracted with TERRA and an alternative pipeline. Our global model includes three orbital equations (circular case). }
         \label{Table:percentilesmultidata}
         \centering
   \begin{tabular}{cccc}
            \hline
            \noalign{\smallskip}
            Jump parameter     &  \multicolumn{2}{c}{Best-fit value}  \\
                               & TERRA & alternative pipeline\\
            \noalign{\smallskip}
            \hline
            \noalign{\smallskip}
            \textbf{Stellar activity GP model} & &\\
            \noalign{\smallskip}
            $h$ [m$\,s^{-1}$] & 2.9$^{+0.4}_{-0.3}$ &  3.1$^{+0.5}_{-0.3}$ \\ 
            \noalign{\smallskip}
            $\lambda$ [days] & 40.0$^{+10.4}_{-9.0}$  & 43.5$^{+9.8}_{-8.9}$ \\ 
            \noalign{\smallskip}
            $w$ & 0.18$^{+0.11}_{-0.04}$ & 0.26$\pm0.07$ \\ 
            \noalign{\smallskip}
            $\theta$ [days] & 40.4$^{+1.1}_{-1.9}$ & 39.8$^{+1.1}_{-0.9}$ \\ 
            \noalign{\smallskip}
            \hline
            \noalign{\smallskip}
            \textbf{Uncorrelated jitter} & &\\
            \noalign{\smallskip}
            $\sigma_{\rm jit, HARPS-N}$ [m$\,s^{-1}$] & 1.4$\pm$0.3 & 0.6$\pm$0.4 \\ 
            \noalign{\smallskip}
            $\sigma_{\rm jit, HARPS-pre}$ [m$\,s^{-1}$] & 2.2$\pm$0.5 & 2.5$\pm$0.5 \\ 
            \noalign{\smallskip}
            $\sigma_{\rm jit, HARPS-post}$ [m$\,s^{-1}$] & 2.0$\pm$0.6 & 1.6$\pm$0.6 \\ 
            \noalign{\smallskip}
            \hline
            \noalign{\smallskip}
            \textbf{RV offset} & &\\
            \noalign{\smallskip}
            $\gamma_{\rm HARPS-N}$ [m$\,s^{-1}$] & -0.06$^{+0.54}_{-0.57}$ & 30149.2$^{+0.6}_{-0.7}$ \\ 
            \noalign{\smallskip}
            $\gamma_{\rm HARPS-pre}$ [m$\,s^{-1}$] & 0.20$^{+0.75}_{-0.79}$ & 30480.2$^{+0.8}_{-0.9}$ \\ 
            \noalign{\smallskip}
            $\gamma_{\rm HARPS-post}$ [m$\,s^{-1}$] & 0.002$^{+0.670}_{-0.688}$  & 30479.2$\pm0.8$ \\ 
            \noalign{\smallskip}
            \hline
            \noalign{\smallskip}
            \textbf{Planetary orbital parameters} & &\\
            \noalign{\smallskip}
            $K_{\rm b}$ [m$\,s^{-1}$] & 2.7$\pm$0.4 & 2.9$\pm$0.4 \\ 
            \noalign{\smallskip}
            $P_{\rm b}$ [days] & 10.05454$\pm$0.00003 & 10.05454$\pm$0.00003 \\ 
            \noalign{\smallskip}
            $T_{\rm 0,b}$ [BJD-2\,400\,000] & 56813.42022$\pm$0.00095 & 56813.42025$\pm$0.00095 \\ 
            \noalign{\smallskip}
            $K_{\rm c}$ [m$\,s^{-1}$] & 0.95$\pm$0.37 & 0.98$\pm$0.34 \\ 
            \noalign{\smallskip}
            $P_{\rm c}$ [days] & 24.64638$\pm$0.00017 & 24.64638$\pm$0.00018 \\ 
            \noalign{\smallskip}
            $T_{\rm 0,c}$ [BJD-2\,400\,000] & 56812.2777$\pm$0.0026 & 56812.2778$\pm$0.0026\\ 
            \noalign{\smallskip}
            $K_{\rm d}$ [m$\,s^{-1}$] & 0.29$^{+0.34}_{-0.18}$ [<0.43 (68.3$^{th}$ perc.)] & 0.31$^{+0.35}_{-0.20}$ [<0.47 (68.3$^{th}$ perc.)] \\ 
            \noalign{\smallskip}
            $P_{\rm d}$ [days] & 44.55764$\pm$0.00042 & 44.55766$\pm$0.00043 \\ 
            \noalign{\smallskip}
            $T_{\rm 0,d}$ [BJD-2\,400\,000] & 56826.2248$\pm$0.0037 & 56826.2247$\pm$0.0038 \\ 
            \noalign{\smallskip}
            \hline
            \noalign{\smallskip}
            \noalign{\smallskip}
            \textbf{Planetary radii} \tablefoottext{a}  \\
            $R_{\rm p, b}$ (R$_{\rm \oplus}$) & \multicolumn{2}{c}{2.29$\pm$0.23} \\ 
            \noalign{\smallskip} 
            $R_{\rm p, c}$ (R$_{\rm \oplus}$) & \multicolumn{2}{c}{1.77$\pm$0.18} \\ 
            \noalign{\smallskip} 
            $R_{\rm p, d}$ (R$_{\rm \oplus}$) & \multicolumn{2}{c}{1.65$\pm$0.17} \\ 
            \noalign{\smallskip} 
            \textbf{Quantities derived from RVs} \tablefoottext{b} \\
            $M_{\rm p, b}$ ($\mearth$) & 6.6$\pm1.1$ & 7.0$\pm$1.0 \\
            \noalign{\smallskip} 
            $M_{\rm p, c}$ ($\mearth$) & 3.1$^{+1.3}_{-1.2}$ & 3.2$^{+1.2}_{-1.1}$ \\ 
            \noalign{\smallskip} 
            $M_{\rm p, d}$ ($\mearth$) & 1.2$^{+1.4}_{-0.7}$ [<1.8 (68.3$^{th}$ perc.)] & 1.3$^{+1.5}_{-0.8}$ [<1.9 (68.3$^{th}$ perc.)] \\ 
            \noalign{\smallskip} 
                                       & 2.7$^{+1.2}_{-0.8}$ (see Sect. \ref{Section:simul1}) & \\
            \noalign{\smallskip}                          

            $\rho_{\rm p, b}$ [g$\,cm^{-3}$] & 3.0$^{+1.3}_{-0.9}$ & 3.2$^{+1.3}_{-0.9}$ \\ 
            \noalign{\smallskip} 
            $\rho_{\rm p, c}$ [g$\,cm^{-3}$] & 3.1$^{+1.9}_{-1.3}$ & 3.2$^{+1.8}_{-1.3}$ \\  
            \noalign{\smallskip} 
            $\rho_{\rm p, d}$ [g$\,cm^{-3}$] & 1.6$^{+2.1}_{-1.0}$ [<2.4 (68.3$^{th}$ perc.)] & 1.6$^{+2.1}_{-1.0}$ \\ 
            \noalign{\smallskip}
                                             & 3.1$_{\rm -1.2}^{\rm +2.0}$ (see Sect. \ref{Section:simul1}) & \\
            \noalign{\smallskip}
            $a_{\rm p, b}$ [AU] & 0.0777$^{+0.0024}_{-0.0026}$ & \\ 
            \noalign{\smallskip} 
            $a_{\rm p, c}$ [AU] & 0.1413$^{+0.0044}_{-0.0047}$ & \\  
            \noalign{\smallskip} 
            $a_{\rm p, d}$ [AU] & 0.2097$^{+0.0065}_{-0.0070}$  & \\ 
            \noalign{\smallskip}
            \noalign{\smallskip}
            \hline
     \end{tabular}    
     \tablefoot{
     \tablefoottext{a}{Radii are derived using our estimate for the stellar radius, R=0.060$\pm$0.06 $\rsun$, and the ratios $R_{\rm planet}/R_{\rm star}$ derived by \cite{beichman16} from \textit{K2} and \textit{Spitzer} data.} 
     \tablefoottext{b}{Derived quantities from the posterior distributions. We used the following equations (assuming  $M_{\rm s}+m_{\rm p} \cong M_{\rm s}$): $m_{\rm p}\sini \cong$ ($K_{\rm p} \cdot M_{\rm s}^{\frac{2}{3}} \cdot \sqrt{1-e^{2}} \cdot P_{\rm p}^{\frac{1}{3}}) / (2\pi G)^{\frac{1}{3}}$; $a \cong [(M_{\rm s}\cdot G)^{\frac{1}{3}}\cdot P_{\rm p}^{\frac{2}{3}}]/(2\pi)^{\frac{2}{3}} $, where $G$ is the gravitational constant.}}
\end{table*}

\begin{figure*}
\centering
\includegraphics[width=0.9\textwidth]{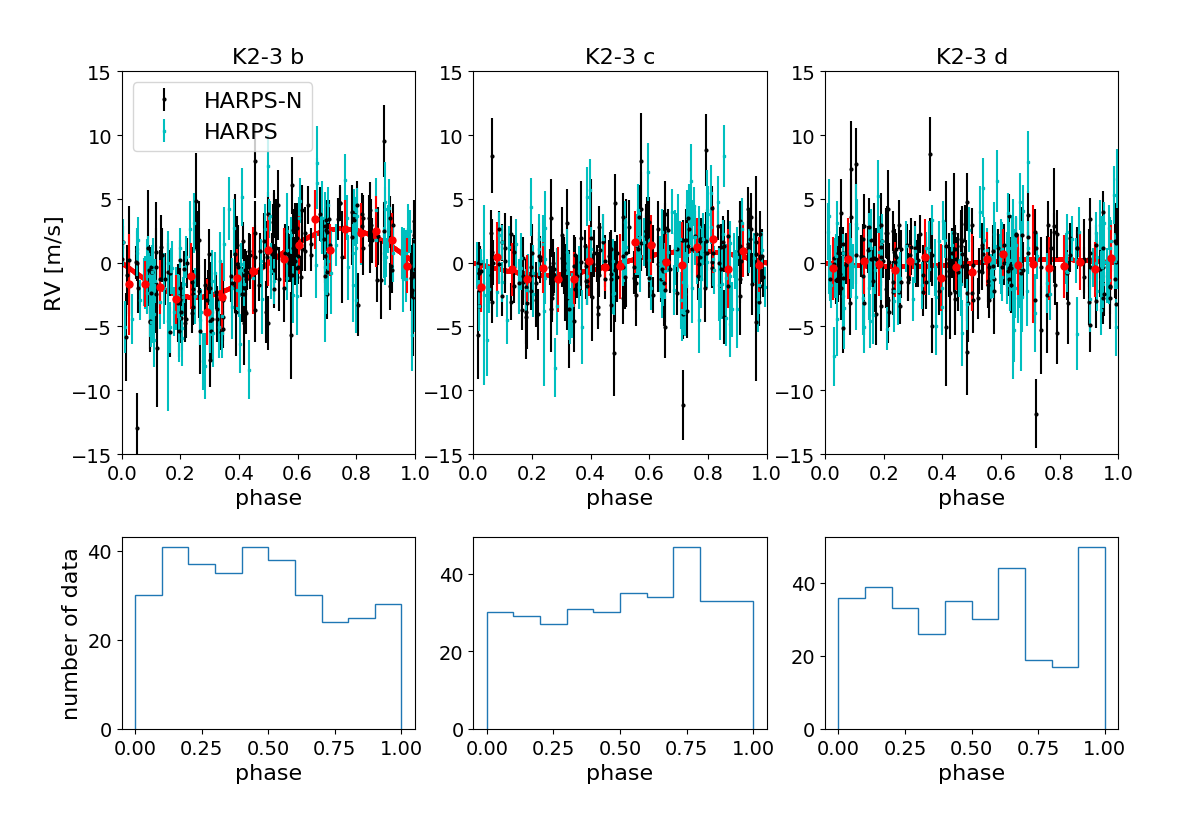}
\caption{\label{fig:foldedrv_terra_qp} \textit{First row:} TERRA RV residuals, after removing our best-fit stellar component, phase-folded to the three planetary solutions found for the quasi-periodic GP model (represented by a red curve). \textit{Second row:} histograms of the number of data divided in bins of phase, showing a fairly uniform coverage for each planet.}
\end{figure*}

\begin{figure}
\centering
\includegraphics[width=0.5\textwidth]{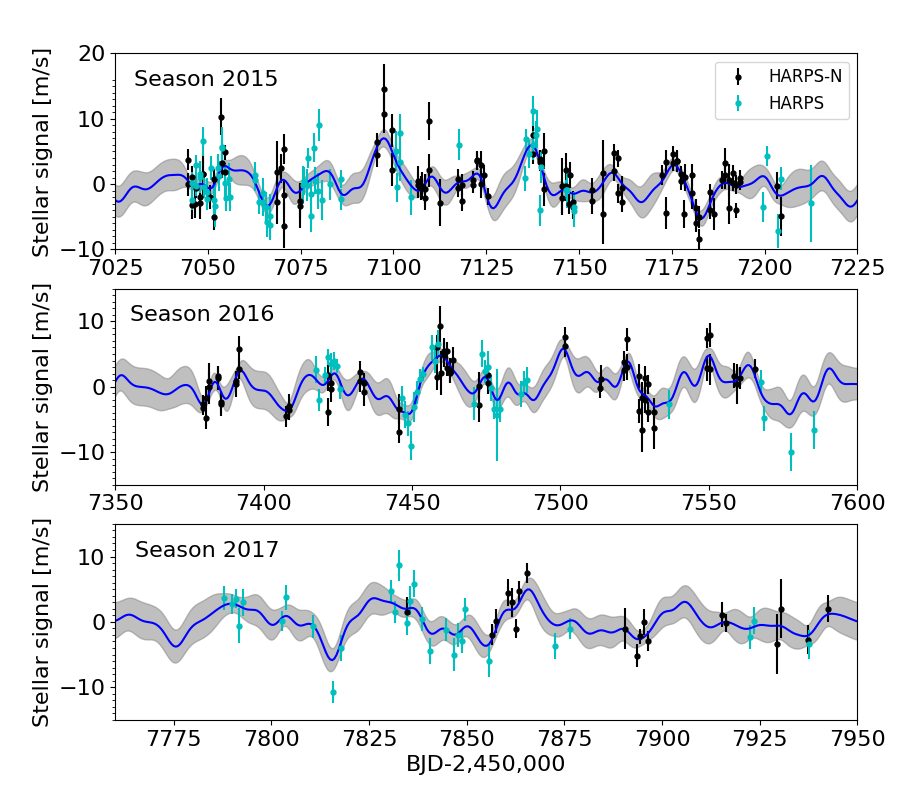}
\caption{\label{fig:stellarnoise_terra_qp} The stellar signal contribution to the TERRA RV times series, for each observing season, as fitted with our GP quasi-periodic model (Table \ref{Table:percentilesmultidata}). The blue line is the best-fit curve, while the shaded grey area represents the $\pm1\sigma$ confidence interval.}
\end{figure}

\begin{figure}
\centering
\includegraphics[width=0.5\textwidth]{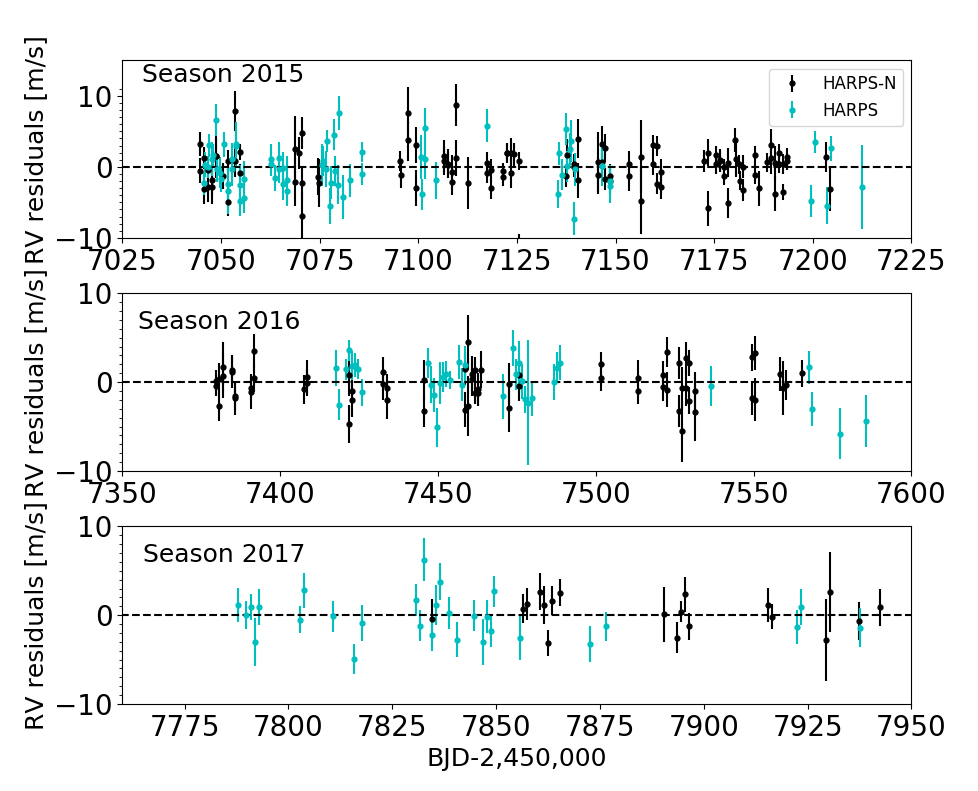}
\caption{\label{fig:residualsgp_qp_terra} TERRA RV residuals, after removing the planetary and stellar activity signals, as fitted within a GP quasi-periodic framework (Table \ref{Table:percentilesmultidata}).}
\end{figure}


\section{Assessing the reliability of the derived planetary masses}
\label{Section:simul}
The results of the analysis presented in Table \ref{Table:percentilesmultidata} show that while the Doppler signal of the innermost planet is retrieved with a significance of $\sim$7$\sigma$, the Doppler signal of K2-3\,c is retrieved with a significance lower than 3$\sigma$, and K2-3\,d is undetected. With the present dataset and the adopted model, we cannot determine a bulk density of K2-3\,c accurately enough to put reliable constraints on its possible composition, and we can only place an upper limit on the bulk density of K2-3\,d. This impacts our understanding of the formation and evolutionary scenarios which led to the observed system architecture, and makes inferences about the habitability of K2-3\,d especially challenging. 

Why is the characterization of this system so challenging, despite the large dataset we have collected with two world-class spectrographs? Can we rule out a rocky composition for K2-3\,d? To provide answers to these questions we have performed simulations to investigate the impact of the observing sampling, stellar activity, and internal RV uncertainties on our ability to retrieve masses for the K2-3 planets. 

\subsection{Simulations using real epochs}
\label{Section:simul1}
We ran a first set of simulations based on the real epochs of our observations with the objective of improving the estimate of K2-3\,d's mass, for which we could provide only an upper limit, and for assessing the robustness of our result for K2-3\,c's mass, despite its low significance. First, we adopted our best-fit values of the GP hyperparameters (TERRA dataset) to generate the stellar activity RV signal. We then injected planetary signals (circular orbits) with semi-amplitudes $K_{\rm b}$=2.7 $\ms$, $K_{\rm c}$=0.95 $\ms$, and $K_{\rm d}$=1 $\ms$, and the known periods and times of transit (Table \ref{Table:percentilesmultidata}) into the simulated activity signal. We refer to the dataset built in this way as the exact solution. While the simulated Doppler semi-amplitudes for K2-3\,b and K2-3\,c are those we have derived from real data, for K2-3\,d we have assumed a value corresponding approximately to a purely rocky composition, given the measured radius\footnote{$K$=1 \ms corresponds to $M_{\rm p}$=4 M$_{\rm \oplus}$.}. 

We then created $N$=156 mock RV time series\footnote{The number of simulations is as large as possible given the computational expense of a GP analysis on each simulated dataset (up to$\sim$10 hours each).}. Each dataset was obtained by randomly drawing RV values normally distributed around the exact solution, where $\sigma^{2}_{\rm RV}(t)+\sigma^{2}_{\rm jit, instr}$ was used as the variance of the normal distributions at each epoch. We analysed each synthetic dataset following the same procedure as for the real data (Section \ref{Section:gpn+s}) and recorded the best-fit values of the free parameters once the MCMC chains reached convergence. 

Results from these simulations depend on the actual properties of the stellar activity observed during our campaign. More complex simulations to explore in detail the effects of the stellar activity could also be carried out, where each hyperparameter is randomly drawn within the uncertainties while keeping the others fixed. However, such simulations, which would require a large number of mock datasets (i.e. thousands) and a correspondingly huge amount of computational time, are beyond the scope of this paper. Thus, our simulations do not explore the possibility that the non-detection is due to the proximity of the stellar rotation period $\theta$ to the orbital period of K2-3\,d, which could be a limiting factor \citep{vander16}. We note that the amplitude \textit{h} of the stellar activity term is precisely known ($\sim10\sigma$, Tab. \ref{Table:percentilesmultidata}), so we expect that drawing values from the posterior distribution of this hyperparameter would not significantly change the results of our simulations. 
          
\textit{Analysis framework}. For each \textit{i}$^{\rm th}$ simulated dataset we derive a posterior distribution for the Doppler semi-amplitudes of planets c and d that we call $K_{\rm p, i}$, where $p$=(c, d). Each semi-amplitude is characterized by a median value $K_{\rm p,i}^{\rm med}$ and upper and lower uncertainties $\sigma_{\rm p,i}^{\rm +}$ and $\sigma_{\rm p,i}^{\rm -}$ (as derived from the $16^{\rm th}$ and $84^{\rm th}$ percentiles). We then calculate the median recovered semi-amplitude $K_{\rm p,N}^{\rm med}$ of all the $K_{\rm p,i}^{\rm med}$. We compare the median recovered semi-amplitude $K_{\rm p,N}^{\rm med}$ with the injected value $K_{\rm p,inj}$ in order to draw conclusions about the results obtained for the real RV dataset.  

For K2-3\,d, we define the ratio $r_{\rm d,i}$=($K_{\rm d,inj}$-$K_{\rm d,i}^{\rm med}$)/$\sigma_{\rm d,i}^{\rm +}$ to measure the discrepancy between the best-fit estimate and the injected value in units of $\sigma_{\rm d,i}^{\rm +}$. The term (K$_{\rm d,inj}$-$K_{\rm d,i}^{\rm med}$) is weighted by $\sigma_{\rm d,i}^{\rm +}$ in order to take the skewness of each posterior distribution into account. By averaging $r_{\rm d,i}$ over the number, $N$, of simulated datasets we get the metric $\overline{r}_{\rm d}$, that we propose as a way to correct the measured semi-amplitude K$_{\rm d,meas}$ using the equation 
\begin{eqnarray} \label{eq:2}
K_{\rm d,real}=K_{\rm d,meas}+ \overline{r}_{\rm d}\cdot\sigma_{\rm d,meas}^{\rm +},
\end{eqnarray}
where $K_{\rm d,meas}$ and $\sigma_{\rm d,meas}^{\rm +}$  come from our best-fit solution (Tab.\ref{Table:percentilesmultidata}).

As an alternative approach, for each marginal distribution $K_{\rm d,i}$ we calculate the percentile corresponding to the position of $K_{\rm d,inj}$. We use the median over $N$ of these percentiles to derive an estimate for $K_{\rm d,real}$ from the posterior distribution obtained for the real dataset.

\textit{Results for K2-3\,c}. 
The distribution of the median values $K_{\rm c,i}^{\rm med}$ is shown in Fig. \ref{fig:kc_simul_dist_50}. The median of this distribution is $K_{\rm c,N}^{\rm med}$=0.96$^{+0.27}_{-0.22}$ \ms, where the uncertainties represent the 16$^{\rm th}$ and 84$^{\rm th}$ percentiles. Looking at the values for $\sigma_{\rm c,i}^{\rm +}$ and $\sigma_{\rm c,i}^{\rm -}$ over all the marginal posteriors $K_{\rm c, i}$, we note that <( $\sigma_{\rm c,i}^{\rm +}$-$\sigma_{\rm c,i}^{\rm -}$)> = 0.01 \ms, indicating that the distributions are normal-shaped, and their average is <$\sigma_{\rm c,i}^{\rm +}$> = < $\sigma_{\rm c,i}^{\rm -}$> = 0.33 \ms. These results show that the injected signal $K_{\rm c,inj}$=0.95 $\ms$ is well recovered, and indicate that our estimate of $K_{\rm c}$ from the real dataset is reliable, despite our detection having a significance < 3-$\sigma$. 

\textit{Results for K2-3\,d}. 
A sample of the $N$ posterior distributions $K_{\rm d,i}$ is shown in Fig. \ref{fig:kd_simul_dist}. The distribution of $K_{\rm d,i}^{\rm med}$ is shown in Fig. \ref{fig:kd_simul_dist_50}. The median of this distribution is $K_{\rm d,N}^{\rm med}$=0.54$^{+0.18}_{-0.14}$ \ms, while we get <$\sigma_{\rm d,i}^{\rm +}$>=0.47 \ms and <$\sigma_{\rm d,i}^{\rm -}$>=0.35 \ms for the 68.3$\%$ confidence interval of each distribution. Therefore, the semi-amplitude $K_{\rm d}$ is generally underestimated with respect to $K_{\rm d,inj}$. This result is suggestive when compared to what we get for $K_{\rm c}$. As shown in Fig. \ref{fig:foldedrv_terra_qp}, the phase coverage of our data is uniform for planet K2-3\,c, and fairly uniform for planet K2-3\,d. Then, we would expect two signals with an equal semi-amplitude (here $K_{\rm c,inj}$ $\simeq$ $K_{\rm d,inj}$) to be recovered with similar significance in absence of other hampering factors.
For the metric $\overline{r}_{\rm d}$ we find $\overline{r}_{\rm d}$=0.99 $\pm$0.04, where the error is calculated as RMS[r$_{\rm d,i}$]/$\sqrt[]{N}$. We use this result and our best-fit value for $K_{\rm d,meas}$=0.29$^{+0.34}_{-0.18}$ to draw $N$=10\,000 random values for $\overline{r}_{\rm d}$ and $K_{\rm d,meas}$, obtaining a distribution of \textit{N} samples for K$_{\rm d,real}$ from Eq. \ref{eq:2}. By taking the median of this distribution and the 68.3$\%$ confidence interval we get $K_{\rm d,real}$=0.63$^{+0.32}_{-0.18}$ \ms. This corresponds to $M_{\rm d,real}$=2.7$_{\rm -0.8}^{\rm +1.2}$ $\mearth$ and $\rho_{\rm d,real}$=3.1$_{\rm -1.2}^{\rm +2.0}$ g cm$^{\rm -3}$ for the planet mass and density. 
By using the second approach mentioned above, we find that $K_{\rm d,real}$ corresponds on average at the 83$^{\rm th}$ percentile of the $K_{\rm d}$ posterior distribution for the real dataset, that is $K_{\rm d,real}$=0.62 \ms, which is a result equal to that obtained with the first method.

Our statistical analysis shows that $K_{\rm d,real}$ is <1 \ms with 1$\sigma$ confidence. This in turn suggests that the planets in the K2-3 system may have very similar bulk densities and thus share a similar composition, although this outcome should be taken with caution due to the large uncertainties in the masses and densities for K2-3\,c and K2-3\,d.   

\begin{figure}
\centering
\includegraphics[width=9.5cm]{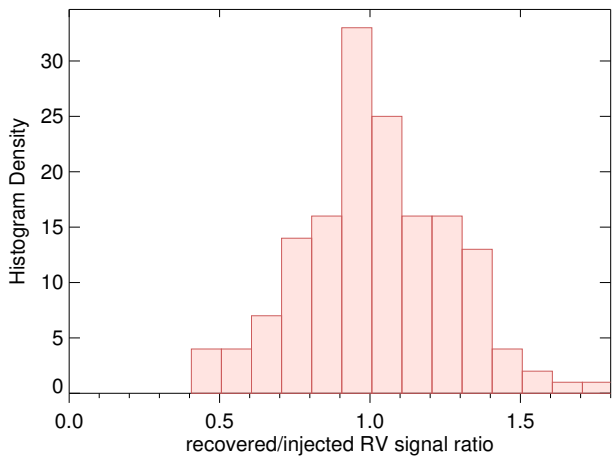}
\caption{\label{fig:kc_simul_dist_50} Distribution of $N$=156 median values for the Doppler semi-amplitude $K_{\rm c}$ of planet K2-3\,c, normalized to the injected value $K_{\rm c, inj}$, as derived from posterior distributions of the $N$ mock RV datasets. This result refers to the case of real observing epochs.}
\end{figure}

\begin{figure}
\centering
\includegraphics[width=9.5cm]{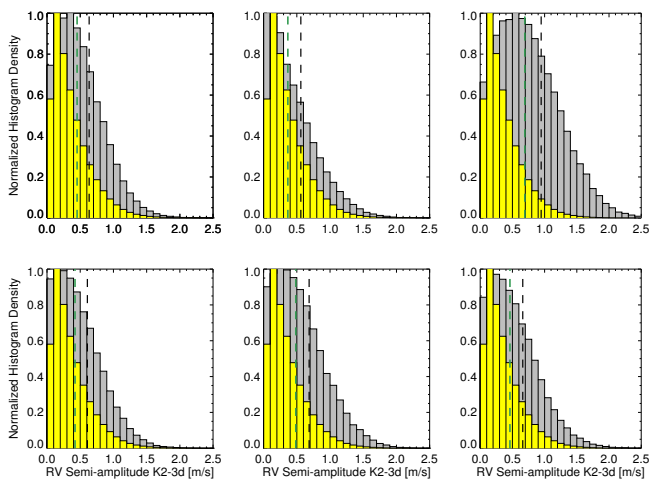}
\caption{\label{fig:kd_simul_dist} Sample of posterior distributions (gray histograms) for the Doppler semi-amplitude $K_{\rm d}$ of planet K2-3\,d obtained from the GP regression analysis of the mock RV datasets described in Sect. \ref{Section:simul1}. Each plot shows the marginal distribution for a single mock dataset, and this is compared to the posterior distribution obtained for the real TERRA RV dataset, represented by the yellow histogram. The vertical dashed lines indicate the 50$th$ (green) and 68.3$th$ (black) percentiles for the posterior distributions of the mock datasets.}
\end{figure}

\begin{figure}
\centering
\includegraphics[width=9.5cm]{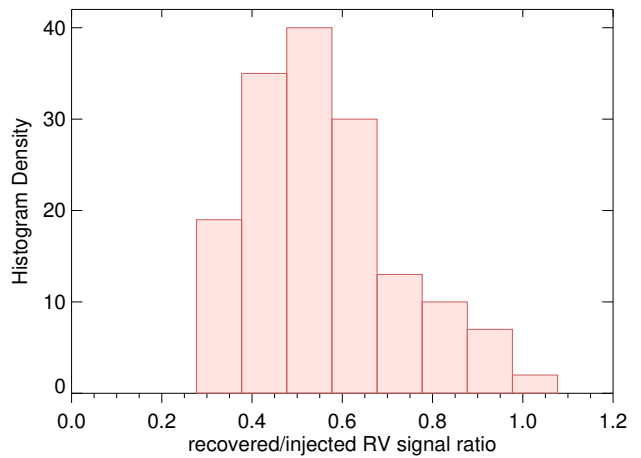}
\caption{\label{fig:kd_simul_dist_50} Distributions of the 50$th$ percentiles for all the posterior distributions of the semi-amplitude $K_{\rm d}$ obtained from 156 mock datasets, normalized to the injected value $K_{\rm d, inj}$. This result refers to the case of real observing epochs.} 
\end{figure}


\subsection{The role of the observing sampling}
\label{Section:simul2}
We devised new simulations to investigate how the detection significance of the signals induced by the planets K2-3\,c and K2-3\,d would change by increasing the number of the RV data collected with HARPS-N and HARPS, still assuming $K_{\rm d}$=1 \ms. We simulated an intensive observing strategy conducted during the third season, which is the one with the lowest number of real data, in a similar way as done for the high-cadence campaign devised to detect Proxima\,b with HARPS \citep{anglada16}. In order to keep our simulation realistic, we did not include mock epochs later than the 2017 observing season because our representation of the RV stellar signal cannot be considered predictive in the far future.

The mock datasets have been created as described in Section \ref{Section:simul1}. We have generated all the epochs suitable for observations during the 2017 observing season in addition to the real ones. We have simulated only one measurement per night avoiding superposition with the epochs corresponding to the real observations. Every random epoch has been selected by placing constraints on the Moon phase illumination and distance from the target (they have to be <90$\%$ and >45$^{\circ}$ , respectively), and on the altitude of the star above the horizon (airmass<1.7). Using these criteria we got 112 and 100 additional new epochs for HARPS-N and HARPS, respectively. We randomly removed 10$\%$ of them at each simulation run to account for bad weather, thus simulating an optimistic scenario for a feasible follow-up\footnote{It must be considered that 10$\%$ refers to the epochs when K2-3 could actually be observed, not to all the nights of the third season.}. The uncertainties $\sigma_{\rm RV}(t)$ of the mock RV data have been randomly drawn from normal distributions, with mean and $\sigma$ equal to the average and RMS values of the HARPS-N and HARPS \textit{post}-upgrade internal errors derived with TERRA. The final mock dataset is obtained by randomly shifting each data point of the exact solution within the error bars by a quantity $\Delta$RV(t) drawn from a normal distribution with mean zero and $\sigma$ equal to $\sqrt[]{\sigma^{\rm 2}_{\rm RV}(t) + \sigma^{\rm 2}_{\rm jit}}$. An example mock dataset is show in Fig. \ref{fig:simulsampl2}.

Our final sample is composed of N=97 mock datasets. Also in this case, we have analysed each simulated dataset within the same GP quasi-periodic framework applied to the real dataset, except for the $\sigma_{\rm jit}$ terms that were not included as free parameters, and we have analysed the outcomes as described in Sect. \ref{Section:simul1}.

\textit{Results for K2-3\,c}. 
The median and the 68.3$\%$ confidence interval of the distribution for the $K_{\rm c,i}$ semi-amplitudes are $K_{\rm c}$=0.96$_{\rm -0.26}^{\rm +0.27}$ \ms. For the upper and lower uncertainties we get <( $\sigma_{\rm c,i}^{\rm +}$-$\sigma_{\rm c,i}^{\rm -}$)> = 0.006 \ms over all the posterior distributions, indicating that they are generally normal-shaped. In addition, <$\sigma_{\rm c,i}^{\rm +}$> = < $\sigma_{\rm c,i}^{\rm -}$> = 0.26 \ms. This result not only confirms that the estimate obtained from real data is robust, but also represents an improvement in the significance of the detection which is now increased to 3.7-$\sigma$.

\textit{Results for K2-3\,d}. 
The median of the $K_{\rm d,i}$ best-fit values is now $K_{\rm d}$=0.58$_{\rm -0.15}^{\rm +0.18}$ \ms, while <$\sigma_{\rm d,i}^{\rm +}$>=0.40 \ms and < $\sigma_{\rm d,i}^{\rm -}$>=-0.34 \ms. This result is not very different from that presented in Sect. \ref{Section:simul1}, and shows that, despite the 190 additional data to the real dataset, the Doppler signal of K2-3\,d is still underestimated and not significantly detected. 

\begin{figure}
\centering
\includegraphics[width=0.5\textwidth]{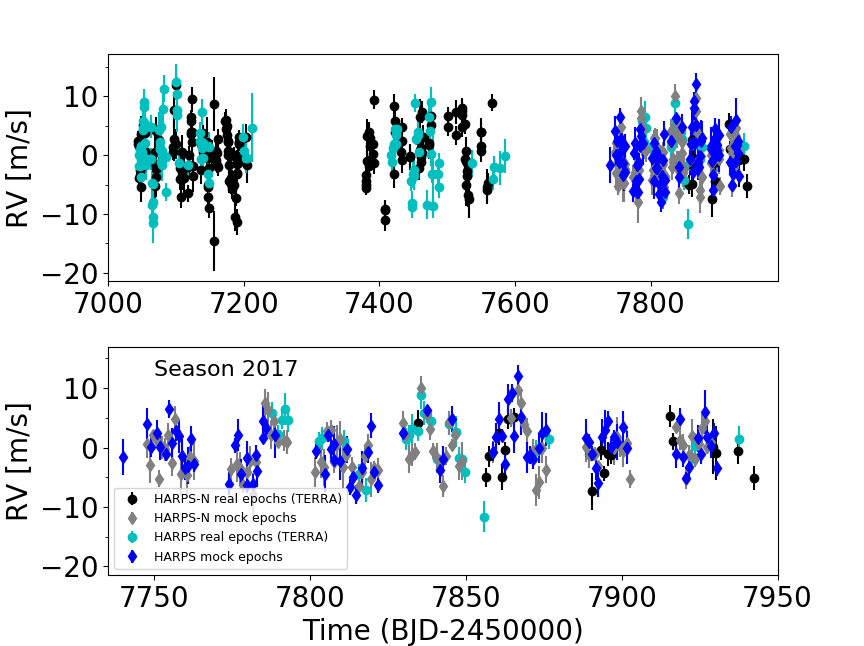}
\caption{\label{fig:simulsampl2} Example of a simulated RV dataset used to explore the effects on the characterization of the K2-3 planets of additional measurements taken over the 2017 season. The upper plot shows the complete mock dataset, while only the third season is shown in the second plot, to better appreciate the intensive simulated sampling.}
\end{figure}


\section{Discussion and conclusions}

This work was focused on deriving the masses and bulk densities of the three planets transiting the nearby M dwarf K2-3, using 329 RV measurements collected with HARPS and HARPS-N over a period of 2.5 years. We found that stellar activity makes a significant contribution to the radial velocity variations over the entire time period. We have also shown that, for the case of K2-3, this can be effectively mitigated using a Gaussian process regression with a quasi-periodic kernel. The results of our global model describe the stellar activity component in a plausible way, and this allowed us to derive a precise and accurate mass estimate for K2-3\,b. We also derive a mass for K2-3\,c with a significance of less than 3$\sigma$. However, using simulations, we demonstrate that our estimate is accurate. Conversely, we do not detect, in our data, the Doppler signal induced by the temperate planet K2-3\,d.

 Figure \ref{fig:massradiusdiag} shows a planetary mass-radius diagram that includes planets for which the mass and radius have been both measured with a relative error better than 30$\%$. Theoretical mass-radius curves for different chemical compositions \citep{zeng13,zeng16} are shown with solid lines. The precision of the radii of the K2-3 planets is mainly limited by that of the stellar radius (all the relative uncertainties are $\sim10\%$; see Table~\ref{Tab:starparam}). One firm outcome of our analysis is that for K2-3\,b an Earth-like composition ($\sim 33\%$ of iron and $\sim67\%$ of silicates) is rejected with high confidence (note that in Fig.~\ref{fig:massradiusdiag} the masses are represented on a logarithmic scale). Concerning K2-3\,c, our mass determination excludes an Earth-like composition with a confidence level of $\sim4\sigma$ (assuming $R_{\rm p}$=1.77\,  $\rearth$). The non-detection of K2-3\,d has been explored in detail through simulations showing that the real Doppler semi-amplitude $K_{\rm d}$ is likely less than 1 \ms and its corresponding mass is $M_{\rm d}=2.7^{+1.2}_{-0.8}$ \, $\mearth$ (Sect.~\ref{Section:simul}). Looking at its position on the mass-radius diagram as derived from our simulations, the interior composition of K2-3\,d would differ from that of the Earth with a confidence level greater than 2$\sigma$ in mass, and $\sim2\sigma$ in radius. We note that planets K2-3\,c and K2-3\,d occupy a region of the mass-radius diagram in which planet occurrence is rare, when only planets with mass and radius measured with a precision better than 30$\%$ are considered.

The corresponding bulk densities of all planets ($\rho_{\rm p}\sim$3 g cm$^{\rm -3}$) show that they may have a very similar composition. 
If further measurements were to confirm our density estimates, excluding rocky compositions for K2-3\,c and K2-3\,d with higher significance, there are two scenarios that may explain the bulk properties of the K2-3 planets: water-poor planets with H/He envelopes or water-worlds.

Several recent studies have investigated the water-poor hypothesis \citep{lopez17,owenwu17,jin17,vaneylen17}. \cite{fulton17} analysed a sample of short-period planetary candidates detected by Kepler (P<100 days), and demonstrated that the distribution of planetary radii is bi-modal: the planet candidates have radii that are predominantly either $\sim1.3$ R$_{\rm \oplus}$ or $\sim2.4$ $R_{\rm \oplus}$, and a gap is observed between 1.5 and 2 R$_{\rm \oplus}$. The evolutionary model of \cite{owenwu17} reproduces the observed bi-modal radius distribution in terms of two populations of planets: those consisting of a bare core, resulting from photoevaporation, and those with twice the core radius, where the size is doubled by a H/He envelopes. The `gap' detected by \cite{fulton17} is actually observed for planets orbiting FGK stars, while the M-dwarf regime was not explored.  However, K2-3 is more similar to the FGK sample, than to a late M dwarf, and the results from Kepler could still be applicable. \footnote{An interesting counterexample is represented by planets orbiting the HZ of the late M-dwarf TRAPPIST-1, which are thought to harbour significant amounts of water \citep{bourrier17}, in particular TRAPPIST-1\,f \citep{quarles17}, showing the diversity of the possible scenarios within the M-dwarf class.} Therefore, K2-3\,b could have a significant volatile envelope, large enough to measurably change its radius with respect to that of a rocky core. The same may be the case for the other two planets in the K2-3 system, but neither of their radius estimates allow us to unambiguously associate them with one of the groups. 

If all planets share the same composition, their densities can be explained by modest primordial hydrogen and helium envelopes atop Earth-like iron and silicate cores. By using the stellar and planet properties derived in this work, and assuming a rocky Earth-like core and a solar composition H/He envelope, we find that K2-3\,b, c, d are best fit with H/He envelopes comprising 0.7$\%$, 0.3$\%$, and 0.4$\%$ of their masses respectively \citep{lopezfort14}. Moreover, using the planetary evolution models of \cite{lopez17} we find that none of the planets in this system are vulnerable to losing significant mass through photo-evaporative atmospheric escape. 

Even though the water-poor scenario has received a great deal of attention, alternatives should also be considered.  Water worlds are planets having massive water envelopes comprising $\geq50\%$ of the planet total mass. Recently, bi-modal radius distributions have been derived for the complete Kepler sample of Q1-Q17 small exoplanet candidates with radius R$_{\rm p}$<4 R$_{\rm \oplus}$  \citep{zeng17b,zeng17c}. These distributions show that the limits and extent of the radius gap depends on the spectral type of the host star. One proposed explanation for the observed bi-modal distributions is the existence of two populations of planets: rocky worlds, with the lowest radii, and water worlds. The two populations likely share the same underlying rocky component by mass, but differ in the presence of a H$_{\rm 2}$O-dominated mantle which is similar to, or slightly more massive than, the rocky component. According to \cite{zeng17c}, the radius gap for an M0 dwarf, like K2-3, is located at 1.6-1.7 R$_{\rm \oplus}$. Therefore K2-3\,b has an observed mass and radius consistent with that expected for a water-world. This could also be the case for K2-3\,c and K2-3\,d.  However, the accuracy of their radii places these planets close to the transition limit that defines the gap, making their water-world membership not highly significant. 

We therefore conclude that all three planets in the K2-3 system are likely sub-Neptunes, defined as small, non-rocky planets which have enough volatiles to measurably change their bulk composition. Both the H/He gas envelope and water-world scenarios are possible, particularly for K2-3\,b. Based on our data, however, we can't rule out that K2-3\,c and K2-3\,d have bare cores of purely rocky/Earth-like composition. 

Within the H/He envelope scenario, planets likely formed within the first $\sim$10 Myr before the dispersion of the gaseous proto-planetary disk. These water-poor planets could have formed in-situ, and their similar masses would suggest similar formation histories \citep{leechiang15,leechiang16}. 
On the other hand, following \cite{ginzburg16} one may expect that the planets, during the cooling phase which follows their formation beyond the snow line (e.g. \citealt{selsis07}), were characterized by an intrinsic luminosity that could blow off, over a billion year timescale, any H/He envelope less than about $\sim5\%$ by mass. Since a H$_{\rm 2}$O-layer has a much higher heat capacity than a H/He envelope, this evolutionary  pathway could result in water-worlds. 

If K2-3\,d is surrounded by a gaseous envelope, with the properties estimated here, this would likely result in surface pressure and temperature too high to support an habitable planet scenario. 
A better characterization of K2-3\,c and K2-3\,d is left to the next generation of high-precision, high-stability spectrographs and to new photometric transit observations. We have shown that an intensive observing sampling over one season with two of the best spectrographs now available would still not have detected a signal with K=1 \ms and  a period P=44.5 days. Assuming that our GP result is a good representation of the stellar activity contribution, this means that the true mass of K2-3\,d is expected to remain unmeasurable even with a dataset of more than 500 RVs, which is currently not possible for a single target and without a collaboration among different teams. Thus, detecting the real signal induced by K2-3\,d is currently very challenging.  K2-3 is, however, an ideal target for characterization studies with the VLT/ESPRESSO spectrograph \citep{pepe14}, or with the NIR spectrographs such as CARMENES \citep{quirrenbach16}, SPIRou \citep{artigau14}, and HPF \citep{mahadevan14}, provided that they reach their design RV precision and assuming, as expected, that RVs extracted from NIR spectra are less affected by stellar activity.
  
\begin{figure*}
\sidecaption
\includegraphics[width=12cm]{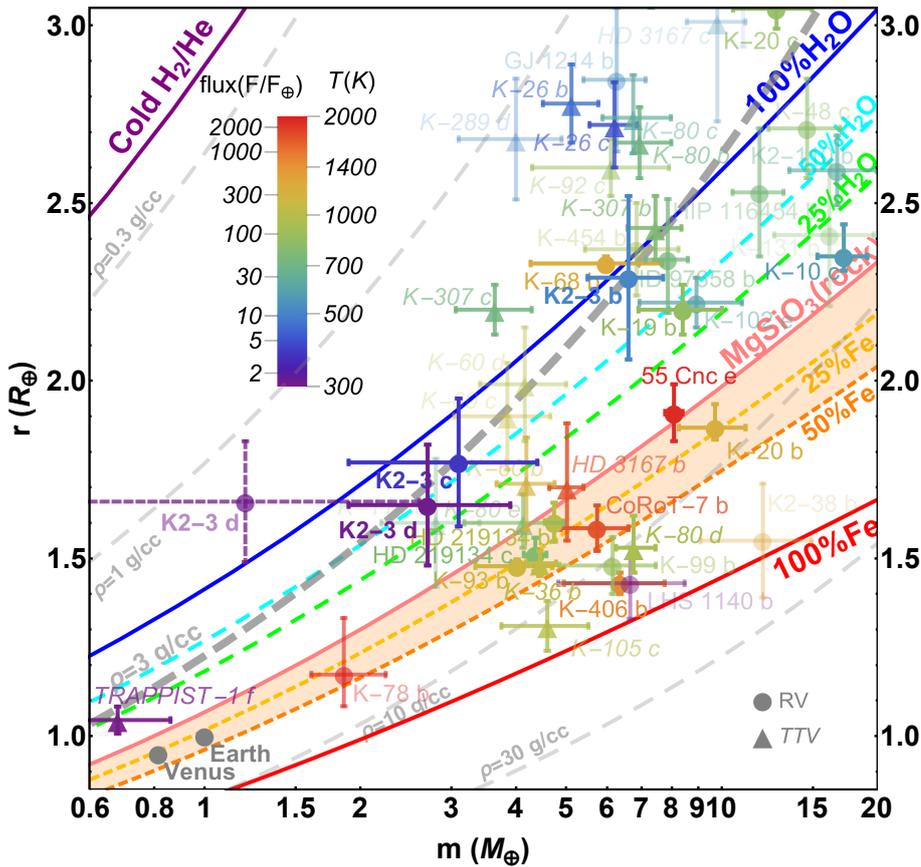}
\caption{Mass-radius diagram for exoplanets for which the mass and radius have been both measured with a relative error better than 30$\%$. The location of the K2-3 planets is emphasized. For K2-3\,d we also plot the mass derived from the GP analysis (shaded point in violet), and the value corrected using the result of the simulations described in Sect. \ref{Section:simul1}. The curve for bulk density $\rho$=3 g cm$^{\rm -3}$ is shown in grey passing through the three positions occupied by the K2-3 planets. The planetary data are taken from the NASA exoplanet archive and updated to 30$^{\rm th}$ August 2017.}
\label{fig:massradiusdiag}
\end{figure*}


\begin{acknowledgements}
    MD acknowledges funding from INAF through the Progetti Premiali funding scheme of the Italian Ministry of Education, University, and Research.
    We used high performance computing resources made available by INAF through the pilot programme \textit{CHIPP} (through the proposal \textit{Precise planetary mass determination in radial velocity data collected with the HARPS and HARPS-N spectrographs: facing the challenges posed by the time sampling and the presence of stellar noise}). We thank especially F. Vitello (INAF-OACt) for his assistance within the CHIPP programme.
    This research has received funding from the European Union Seventh Framework Programme (FP7/2007-2013) under grant Agreement No. 313014 (ETAEARTH).
    Parts of this work have been supported by NASA under grants No. NNX15AC90G and NNX17AB59G issued through the Exoplanets Research Program.
    AVa acknowledges support from the California Institute of
Technology (Caltech)/Jet Propulsion Laboratory (JPL) funded by NASA through the Sagan Fellowship Program executed by the NASA Exoplanet Science Institute.
    PF acknowledges support by Funda\c{c}\~ao para a Ci\^encia e a Tecnologia (FCT) through Investigador FCT contract of reference IF/01037/2013/CP1191/CT0001, and POPH/FSE (EC) by FEDER funding through the program ``Programa Operacional de Factores de Competitividade - COMPETE''. 
    NCS acknowledges support by Funda\c{c}\~ao para a Ci\^encia e a Tecnologia (FCT, Portugal) through the research grant through national funds and by FEDER through COMPETE2020 by grants UID/FIS/04434/2013 $\&$ POCI-01-0145-FEDER-007672 and PTDC/FIS-AST/1526/2014 $\&$ POCI-01-0145-FEDER-016886, as well as through Investigador FCT contract nr. IF/00169/2012/CP0150/CT0002.
\end{acknowledgements}


\bibliographystyle{aa} 
\bibliography{k23.bib} 

\begin{thebibliography}{75}
\expandafter\ifx\csname natexlab\endcsname\relax\def\natexlab#1{#1}\fi

\bibitem[{{Affer} {et~al.}(2016){Affer}, {Micela}, {Damasso}, {Perger},
  {Ribas}, {Su{\'a}rez Mascare{\~n}o}, {Gonz{\'a}lez Hern{\'a}ndez}, {Rebolo},
  {Poretti}, {Maldonado}, {Leto}, {Pagano}, {Scandariato}, {Zanmar Sanchez},
  {Sozzetti}, {Bonomo}, {Malavolta}, {Morales}, {Rosich}, {Bignamini},
  {Gratton}, {Velasco}, {Cenadelli}, {Claudi}, {Cosentino}, {Desidera},
  {Giacobbe}, {Herrero}, {Lafarga}, {Lanza}, {Molinari}, \& {Piotto}}]{affer16}
{Affer}, L., {Micela}, G., {Damasso}, M., {et~al.} 2016, \aap, 593, A117

\bibitem[{{Almenara} {et~al.}(2015){Almenara}, {Astudillo-Defru}, {Bonfils},
  {Forveille}, {Santerne}, {Albrecht}, {Barros}, {Bouchy}, {Delfosse},
  {Demangeon}, {Diaz}, {H{\'e}brard}, {Mayor}, {Neves}, {Rojo}, {Santos}, \&
  {W{\"u}nsche}}]{almenara15}
{Almenara}, J.~M., {Astudillo-Defru}, N., {Bonfils}, X., {et~al.} 2015, \aap,
  581, L7

\bibitem[{{Anglada-Escud{\'e}} {et~al.}(2016){Anglada-Escud{\'e}}, {Amado},
  {Barnes}, {Berdi{\~n}as}, {Butler}, {Coleman}, {de La Cueva}, {Dreizler},
  {Endl}, {Giesers}, {Jeffers}, {Jenkins}, {Jones}, {Kiraga}, {K{\"u}rster},
  {L{\'o}pez-Gonz{\'a}lez}, {Marvin}, {Morales}, {Morin}, {Nelson}, {Ortiz},
  {Ofir}, {Paardekooper}, {Reiners}, {Rodr{\'{\i}}guez},
  {Rodr{\'{\i}}guez-L{\'o}pez}, {Sarmiento}, {Strachan}, {Tsapras}, {Tuomi}, \&
  {Zechmeister}}]{anglada16}
{Anglada-Escud{\'e}}, G., {Amado}, P.~J., {Barnes}, J., {et~al.} 2016, \nat,
  536, 437

\bibitem[{{Anglada-Escud{\'e}} {et~al.}(2013){Anglada-Escud{\'e}}, {Tuomi},
  {Gerlach}, {Barnes}, {Heller}, {Jenkins}, {Wende}, {Vogt}, {Butler},
  {Reiners}, \& {Jones}}]{anglada13}
{Anglada-Escud{\'e}}, G., {Tuomi}, M., {Gerlach}, E., {et~al.} 2013, \aap, 556,
  A126

\bibitem[{Anglada-Escudé \& Butler(2012)}]{anglada12}
Anglada-Escudé, G. \& Butler, R.~P. 2012, The Astrophysical Journal Supplement
  Series, 200, 15

\bibitem[{{Artigau} {et~al.}(2014){Artigau}, {Kouach}, {Donati}, {Doyon},
  {Delfosse}, {Baratchart}, {Lacombe}, {Moutou}, {Rabou}, {Par{\`e}s},
  {Micheau}, {Thibault}, {Reshetov}, {Dubois}, {Hernandez}, {Vall{\'e}e},
  {Wang}, {Dolon}, {Pepe}, {Bouchy}, {Striebig}, {H{\'e}nault}, {Loop},
  {Saddlemyer}, {Barrick}, {Vermeulen}, {Dupieux}, {H{\'e}brard}, {Boisse},
  {Martioli}, {Alencar}, {do Nascimento}, \& {Figueira}}]{artigau14}
{Artigau}, {\'E}., {Kouach}, D., {Donati}, J.-F., {et~al.} 2014, in \procspie,
  Vol. 9147, Ground-based and Airborne Instrumentation for Astronomy V, 914715

\bibitem[{{Astudillo-Defru} {et~al.}(2015){Astudillo-Defru}, {Bonfils},
  {Delfosse}, {S{\'e}gransan}, {Forveille}, {Bouchy}, {Gillon}, {Lovis},
  {Mayor}, {Neves}, {Pepe}, {Perrier}, {Queloz}, {Rojo}, {Santos}, \&
  {Udry}}]{astudillo15}
{Astudillo-Defru}, N., {Bonfils}, X., {Delfosse}, X., {et~al.} 2015, \aap, 575,
  A119

\bibitem[{{Astudillo-Defru} {et~al.}(2017){Astudillo-Defru}, {Forveille},
  {Bonfils}, {S{\'e}gransan}, {Bouchy}, {Delfosse}, {Lovis}, {Mayor}, {Murgas},
  {Pepe}, {Santos}, {Udry}, \& {W{\"u}nsche}}]{astudillo17}
{Astudillo-Defru}, N., {Forveille}, T., {Bonfils}, X., {et~al.} 2017, \aap,
  602, A88

\bibitem[{{Beichman} {et~al.}(2014){Beichman}, {Benneke}, {Knutson}, {Smith},
  {Lagage}, {Dressing}, {Latham}, {Lunine}, {Birkmann}, {Ferruit}, {Giardino},
  {Kempton}, {Carey}, {Krick}, {Deroo}, {Mandell}, {Ressler}, {Shporer},
  {Swain}, {Vasisht}, {Ricker}, {Bouwman}, {Crossfield}, {Greene}, {Howell},
  {Christiansen}, {Ciardi}, {Clampin}, {Greenhouse}, {Sozzetti}, {Goudfrooij},
  {Hines}, {Keyes}, {Lee}, {McCullough}, {Robberto}, {Stansberry}, {Valenti},
  {Rieke}, {Rieke}, {Fortney}, {Bean}, {Kreidberg}, {Ehrenreich}, {Deming},
  {Albert}, {Doyon}, \& {Sing}}]{beichman14}
{Beichman}, C., {Benneke}, B., {Knutson}, H., {et~al.} 2014, \pasp, 126, 1134

\bibitem[{{Beichman} {et~al.}(2016){Beichman}, {Livingston}, {Werner},
  {Gorjian}, {Krick}, {Deck}, {Knutson}, {Wong}, {Petigura}, {Christiansen},
  {Ciardi}, {Greene}, {Schlieder}, {Line}, {Crossfield}, {Howard}, \&
  {Sinukoff}}]{beichman16}
{Beichman}, C., {Livingston}, J., {Werner}, M., {et~al.} 2016, \apj, 822, 39

\bibitem[{{Benatti} {et~al.}(2016){Benatti}, {Claudi}, {Desidera}, {Gratton},
  {Lanza}, {Micela}, {Pagano}, {Piotto}, {Sozzetti}, {Boccato}, {Cosentino},
  {Covino}, {Maggio}, {Molinari}, {Poretti}, {Smareglia}, \& {GAPS
  Team}}]{benatti16}
{Benatti}, S., {Claudi}, R., {Desidera}, S., {et~al.} 2016, in Frontier
  Research in Astrophysics II, held 23-28 May, 2016 in Mondello (Palermo),
  Italy (FRAPWS2016). Online at <A
  href=``href=''>https://pos.sissa.it/cgi-bin/reader/conf.cgi?confid=269</A>,
  id.69, 69

\bibitem[{{Bonfils} {et~al.}(2015){Bonfils}, {Almenara}, {Jocou}, {Wunsche},
  {Kern}, {Delboulb{\'e}}, {Delfosse}, {Feautrier}, {Forveille}, {Gluck},
  {Lafrasse}, {Magnard}, {Maurel}, {Moulin}, {Murgas}, {Rabou}, {Rochat},
  {Roux}, \& {Stadler}}]{bonfils15}
{Bonfils}, X., {Almenara}, J.~M., {Jocou}, L., {et~al.} 2015, in \procspie,
  Vol. 9605, Techniques and Instrumentation for Detection of Exoplanets VII,
  96051L

\bibitem[{{Bonfils} {et~al.}(2017){Bonfils}, {Astudillo-Defru}, {D{\'{\i}}az},
  {Almenara}, {Forveille}, {Bouchy}, {Delfosse}, {Lovis}, {Mayor}, {Murgas},
  {Pepe}, {Santos}, {S{\'e}gransan}, {Udry}, \& {W{\"u}nsche}}]{bonfils17}
{Bonfils}, X., {Astudillo-Defru}, N., {D{\'{\i}}az}, R., {et~al.} 2017, ArXiv
  e-prints [\eprint[arXiv]{1711.06177}]

\bibitem[{{Bonfils} {et~al.}(2013){Bonfils}, {Delfosse}, {Udry}, {Forveille},
  {Mayor}, {Perrier}, {Bouchy}, {Gillon}, {Lovis}, {Pepe}, {Queloz}, {Santos},
  {S{\'e}gransan}, \& {Bertaux}}]{bonfils13}
{Bonfils}, X., {Delfosse}, X., {Udry}, S., {et~al.} 2013, \aap, 549, A109

\bibitem[{Bourrier {et~al.}(2017)Bourrier, de~Wit, Bolmont, Stamenković,
  Wheatley, Burgasser, Delrez, Demory, Ehrenreich, Gillon, Jehin, Leconte,
  Lederer, Lewis, Triaud, \& Grootel}]{bourrier17}
Bourrier, V., de~Wit, J., Bolmont, E., {et~al.} 2017, The Astronomical Journal,
  154, 121

\bibitem[{{Cloutier} {et~al.}(2017){Cloutier}, {Astudillo-Defru}, {Doyon},
  {Bonfils}, {Almenara}, {Benneke}, {Bouchy}, {Delfosse}, {Ehrenreich},
  {Forveille}, {Lovis}, {Mayor}, {Menou}, {Murgas}, {Pepe}, {Rowe}, {Santos},
  {Udry}, \& {W{\"u}nsche}}]{cloutier17}
{Cloutier}, R., {Astudillo-Defru}, N., {Doyon}, R., {et~al.} 2017, \aap, 608,
  A35

\bibitem[{{Cosentino} {et~al.}(2014){Cosentino}, {Lovis}, {Pepe}, {Cameron},
  {Latham}, {Molinari}, {Udry}, {Bezawada}, {Buchschacher}, {Figueira},
  {Fleury}, {Ghedina}, {Glenday}, {Gonzalez}, {Guerra}, {Henry}, {Hughes},
  {Maire}, {Motalebi}, \& {Phillips}}]{cosentino14}
{Cosentino}, R., {Lovis}, C., {Pepe}, F., {et~al.} 2014, in \procspie, Vol.
  9147, Ground-based and Airborne Instrumentation for Astronomy V, 91478C

\bibitem[{Crossfield {et~al.}(2015)Crossfield, Petigura, Schlieder, Howard,
  Fulton, Aller, Ciardi, Lépine, Barclay, de~Pater, de~Kleer, Quintana,
  Christiansen, Schlafly, Kaltenegger, Crepp, Henning, Obermeier, Deacon,
  Weiss, Isaacson, Hansen, Liu, Greene, Howell, Barman, \&
  Mordasini}]{crossfield15}
Crossfield, I. J.~M., Petigura, E., Schlieder, J.~E., {et~al.} 2015, The
  Astrophysical Journal, 804, 10

\bibitem[{{Dai} {et~al.}(2016){Dai}, {Winn}, {Albrecht}, {Arriagada},
  {Bieryla}, {Butler}, {Crane}, {Hirano}, {Johnson}, {Kiilerich}, {Latham},
  {Narita}, {Nowak}, {Palle}, {Ribas}, {Rogers}, {Sanchis-Ojeda}, {Shectman},
  {Teske}, {Thompson}, {Van Eylen}, {Vanderburg}, {Wittenmyer}, \&
  {Yu}}]{dai16}
{Dai}, F., {Winn}, J.~N., {Albrecht}, S., {et~al.} 2016, \apj, 823, 115

\bibitem[{{Damasso} \& {Del Sordo}(2017)}]{damasso17}
{Damasso}, M. \& {Del Sordo}, F. 2017, \aap, 599, A126

\bibitem[{{Delfosse} {et~al.}(2013{\natexlab{a}}){Delfosse}, {Bonfils},
  {Forveille}, {Udry}, {Mayor}, {Bouchy}, {Gillon}, {Lovis}, {Neves}, {Pepe},
  {Perrier}, {Queloz}, {Santos}, \& {S{\'e}gransan}}]{delfosse13b}
{Delfosse}, X., {Bonfils}, X., {Forveille}, T., {et~al.} 2013{\natexlab{a}},
  \aap, 553, A8

\bibitem[{{Delfosse} {et~al.}(2013{\natexlab{b}}){Delfosse}, {Donati},
  {Kouach}, {H{\'e}brard}, {Doyon}, {Artigau}, {Bouchy}, {Boisse}, {Brun},
  {Hennebelle}, {Widemann}, {Bouvier}, {Bonfils}, {Morin}, {Moutou}, {Pepe},
  {Udry}, {do Nascimento}, {Alencar}, {Castilho}, {Martioli}, {Wang},
  {Figueira}, \& {Santos}}]{delfosse13}
{Delfosse}, X., {Donati}, J.-F., {Kouach}, D., {et~al.} 2013{\natexlab{b}}, in
  SF2A-2013: Proceedings of the Annual meeting of the French Society of
  Astronomy and Astrophysics, ed. L.~{Cambresy}, F.~{Martins}, E.~{Nuss}, \&
  A.~{Palacios}, 497--508

\bibitem[{{Dittmann} {et~al.}(2017){Dittmann}, {Irwin}, {Charbonneau},
  {Bonfils}, {Astudillo-Defru}, {Haywood}, {Berta-Thompson}, {Newton},
  {Rodriguez}, {Winters}, {Tan}, {Almenara}, {Bouchy}, {Delfosse}, {Forveille},
  {Lovis}, {Murgas}, {Pepe}, {Santos}, {Udry}, {W{\"u}nsche}, {Esquerdo},
  {Latham}, \& {Dressing}}]{dittman17}
{Dittmann}, J.~A., {Irwin}, J.~M., {Charbonneau}, D., {et~al.} 2017, \nat, 544,
  333

\bibitem[{{Dressing} \& {Charbonneau}(2013)}]{dressing13}
{Dressing}, C.~D. \& {Charbonneau}, D. 2013, \apj, 767, 95

\bibitem[{{Dressing} \& {Charbonneau}(2015)}]{dressing15}
{Dressing}, C.~D. \& {Charbonneau}, D. 2015, \apj, 807, 45

\bibitem[{Dressing {et~al.}(2017)Dressing, Newton, Schlieder, Charbonneau,
  Knutson, Vanderburg, \& Sinukoff}]{dressing17a}
Dressing, C.~D., Newton, E.~R., Schlieder, J.~E., {et~al.} 2017, The
  Astrophysical Journal, 836, 167

\bibitem[{{Dressing} {et~al.}(2017){Dressing}, {Vanderburg}, {Schlieder},
  {Crossfield}, {Knutson}, {Newton}, {Ciardi}, {Fulton}, {Gonzales}, {Howard},
  {Isaacson}, {Livingston}, {Petigura}, {Sinukoff}, {Everett}, {Horch}, \&
  {Howell}}]{dressing17b}
{Dressing}, C.~D., {Vanderburg}, A., {Schlieder}, J.~E., {et~al.} 2017, \aj,
  154, 207

\bibitem[{{Dumusque} {et~al.}(2017){Dumusque}, {Borsa}, {Damasso},
  {D{\'{\i}}az}, {Gregory}, {Hara}, {Hatzes}, {Rajpaul}, {Tuomi}, {Aigrain},
  {Anglada-Escud{\'e}}, {Bonomo}, {Bou{\'e}}, {Dauvergne}, {Frustagli},
  {Giacobbe}, {Haywood}, {Jones}, {Laskar}, {Pinamonti}, {Poretti}, {Rainer},
  {S{\'e}gransan}, {Sozzetti}, \& {Udry}}]{dumusque17}
{Dumusque}, X., {Borsa}, F., {Damasso}, M., {et~al.} 2017, \aap, 598, A133

\bibitem[{{Eastman} {et~al.}(2013){Eastman}, {Gaudi}, \& {Agol}}]{eastman13}
{Eastman}, J., {Gaudi}, B.~S., \& {Agol}, E. 2013, \pasp, 125, 83

\bibitem[{{Feroz} \& {Hobson}(2014)}]{feroz14}
{Feroz}, F. \& {Hobson}, M.~P. 2014, \mnras, 437, 3540

\bibitem[{{Fortier} {et~al.}(2014){Fortier}, {Beck}, {Benz}, {Broeg}, {Cessa},
  {Ehrenreich}, \& {Thomas}}]{fortier14}
{Fortier}, A., {Beck}, T., {Benz}, W., {et~al.} 2014, in \procspie, Vol. 9143,
  Space Telescopes and Instrumentation 2014: Optical, Infrared, and Millimeter
  Wave, 91432J

\bibitem[{{Fulton} {et~al.}(2017){Fulton}, {Petigura}, {Howard}, {Isaacson},
  {Marcy}, {Cargile}, {Hebb}, {Weiss}, {Johnson}, {Morton}, {Sinukoff},
  {Crossfield}, \& {Hirsch}}]{fulton17}
{Fulton}, B.~J., {Petigura}, E.~A., {Howard}, A.~W., {et~al.} 2017, \aj, 154,
  109

\bibitem[{{Giles} {et~al.}(2017){Giles}, {Collier Cameron}, \&
  {Haywood}}]{giles17}
{Giles}, H.~A.~C., {Collier Cameron}, A., \& {Haywood}, R.~D. 2017, \mnras,
  472, 1618

\bibitem[{{Gillon} {et~al.}(2017){Gillon}, {Triaud}, {Demory}, {Jehin}, {Agol},
  {Deck}, {Lederer}, {de Wit}, {Burdanov}, {Ingalls}, {Bolmont}, {Leconte},
  {Raymond}, {Selsis}, {Turbet}, {Barkaoui}, {Burgasser}, {Burleigh}, {Carey},
  {Chaushev}, {Copperwheat}, {Delrez}, {Fernandes}, {Holdsworth}, {Kotze}, {Van
  Grootel}, {Almleaky}, {Benkhaldoun}, {Magain}, \& {Queloz}}]{gillon17}
{Gillon}, M., {Triaud}, A.~H.~M.~J., {Demory}, B.-O., {et~al.} 2017, \nat, 542,
  456

\bibitem[{{Ginzburg} {et~al.}(2016){Ginzburg}, {Schlichting}, \&
  {Sari}}]{ginzburg16}
{Ginzburg}, S., {Schlichting}, H.~E., \& {Sari}, R. 2016, \apj, 825, 29

\bibitem[{{Gomes da Silva} {et~al.}(2011){Gomes da Silva}, {Santos}, {Bonfils},
  {Delfosse}, {Forveille}, \& {Udry}}]{gomesdasilva11}
{Gomes da Silva}, J., {Santos}, N.~C., {Bonfils}, X., {et~al.} 2011, \aap, 534,
  A30

\bibitem[{{Haywood} {et~al.}(2014){Haywood}, {Collier Cameron}, {Queloz},
  {Barros}, {Deleuil}, {Fares}, {Gillon}, {Lanza}, {Lovis}, {Moutou}, {Pepe},
  {Pollacco}, {Santerne}, {S{\'e}gransan}, \& {Unruh}}]{haywood14}
{Haywood}, R.~D., {Collier Cameron}, A., {Queloz}, D., {et~al.} 2014, \mnras,
  443, 2517

\bibitem[{{Howell} {et~al.}(2014){Howell}, {Sobeck}, {Haas}, {Still},
  {Barclay}, {Mullally}, {Troeltzsch}, {Aigrain}, {Bryson}, {Caldwell},
  {Chaplin}, {Cochran}, {Huber}, {Marcy}, {Miglio}, {Najita}, {Smith},
  {Twicken}, \& {Fortney}}]{howell14}
{Howell}, S.~B., {Sobeck}, C., {Haas}, M., {et~al.} 2014, \pasp, 126, 398

\bibitem[{{Irwin} {et~al.}(2015){Irwin}, {Berta-Thompson}, {Charbonneau},
  {Dittmann}, {Falco}, {Newton}, \& {Nutzman}}]{irwin15}
{Irwin}, J.~M., {Berta-Thompson}, Z.~K., {Charbonneau}, D., {et~al.} 2015, in
  Cambridge Workshop on Cool Stars, Stellar Systems, and the Sun, Vol.~18, 18th
  Cambridge Workshop on Cool Stars, Stellar Systems, and the Sun, ed. G.~T.
  {van Belle} \& H.~C. {Harris}, 767--772

\bibitem[{{Jin} \& {Mordasini}(2017)}]{jin17}
{Jin}, S. \& {Mordasini}, C. 2017, ArXiv e-prints [\eprint[arXiv]{1706.00251}]

\bibitem[{{Kopparapu} {et~al.}(2013){Kopparapu}, {Ramirez}, {Kasting}, {Eymet},
  {Robinson}, {Mahadevan}, {Terrien}, {Domagal-Goldman}, {Meadows}, \&
  {Deshpande}}]{koppa13}
{Kopparapu}, R.~K., {Ramirez}, R., {Kasting}, J.~F., {et~al.} 2013, \apj, 765,
  131

\bibitem[{{Kopparapu} {et~al.}(2014){Kopparapu}, {Ramirez}, {SchottelKotte},
  {Kasting}, {Domagal-Goldman}, \& {Eymet}}]{koppa14}
{Kopparapu}, R.~K., {Ramirez}, R.~M., {SchottelKotte}, J., {et~al.} 2014,
  \apjl, 787, L29

\bibitem[{{Lee} \& {Chiang}(2015)}]{leechiang15}
{Lee}, E.~J. \& {Chiang}, E. 2015, \apj, 811, 41

\bibitem[{{Lee} \& {Chiang}(2016)}]{leechiang16}
{Lee}, E.~J. \& {Chiang}, E. 2016, \apj, 817, 90

\bibitem[{{Lo Curto} {et~al.}(2015){Lo Curto}, {Pepe}, {Avila}, {Boffin},
  {Bovay}, {Chazelas}, {Coffinet}, {Fleury}, {Hughes}, {Lovis}, {Maire},
  {Manescau}, {Pasquini}, {Rihs}, {Sinclaire}, \& {Udry}}]{locurto15}
{Lo Curto}, G., {Pepe}, F., {Avila}, G., {et~al.} 2015, The Messenger, 162, 9

\bibitem[{{Lopez}(2017)}]{lopez17}
{Lopez}, E.~D. 2017, \mnras, 472, 245

\bibitem[{{Lopez} \& {Fortney}(2014)}]{lopezfort14}
{Lopez}, E.~D. \& {Fortney}, J.~J. 2014, \apj, 792, 1

\bibitem[{{L{\'o}pez-Morales} {et~al.}(2016){L{\'o}pez-Morales}, {Haywood},
  {Coughlin}, {Zeng}, {Buchhave}, {Giles}, {Affer}, {Bonomo}, {Charbonneau},
  {Collier Cameron}, {Consentino}, {Dressing}, {Dumusque}, {Figueira},
  {Fiorenzano}, {Harutyunyan}, {Johnson}, {Latham}, {Lopez}, {Lovis},
  {Malavolta}, {Mayor}, {Micela}, {Molinari}, {Mortier}, {Motalebi},
  {Nascimbeni}, {Pepe}, {Phillips}, {Piotto}, {Pollacco}, {Queloz}, {Rice},
  {Sasselov}, {Segransan}, {Sozzetti}, {Udry}, {Vanderburg}, \&
  {Watson}}]{morales16}
{L{\'o}pez-Morales}, M., {Haywood}, R.~D., {Coughlin}, J.~L., {et~al.} 2016,
  \aj, 152, 204

\bibitem[{{Lovis} \& {Pepe}(2007)}]{lovis07}
{Lovis}, C. \& {Pepe}, F. 2007, \aap, 468, 1115

\bibitem[{{Mahadevan} {et~al.}(2014){Mahadevan}, {Ramsey}, {Terrien},
  {Halverson}, {Roy}, {Hearty}, {Levi}, {Stefansson}, {Robertson}, {Bender},
  {Schwab}, \& {Nelson}}]{mahadevan14}
{Mahadevan}, S., {Ramsey}, L.~W., {Terrien}, R., {et~al.} 2014, in \procspie,
  Vol. 9147, Ground-based and Airborne Instrumentation for Astronomy V, 91471G

\bibitem[{{Malavolta} {et~al.}(2017){Malavolta}, {Borsato}, {Granata},
  {Piotto}, {Lopez}, {Vanderburg}, {Figueira}, {Mortier}, {Nascimbeni},
  {Affer}, {Bonomo}, {Bouchy}, {Buchhave}, {Charbonneau}, {Collier Cameron},
  {Cosentino}, {Dressing}, {Dumusque}, {Fiorenzano}, {Harutyunyan}, {Haywood},
  {Johnson}, {Latham}, {Lopez-Morales}, {Lovis}, {Mayor}, {Micela}, {Molinari},
  {Motalebi}, {Pepe}, {Phillips}, {Pollacco}, {Queloz}, {Rice}, {Sasselov},
  {S{\'e}gransan}, {Sozzetti}, {Udry}, \& {Watson}}]{malavolta17}
{Malavolta}, L., {Borsato}, L., {Granata}, V., {et~al.} 2017, \aj, 153, 224

\bibitem[{{Maldonado} {et~al.}(2015){Maldonado}, {Affer}, {Micela},
  {Scandariato}, {Damasso}, {Stelzer}, {Barbieri}, {Bedin}, {Biazzo},
  {Bignamini}, {Borsa}, {Claudi}, {Covino}, {Desidera}, {Esposito}, {Gratton},
  {Gonz{\'a}lez Hern{\'a}ndez}, {Lanza}, {Maggio}, {Molinari}, {Pagano},
  {Perger}, {Pillitteri}, {Piotto}, {Poretti}, {Prisinzano}, {Rebolo}, {Ribas},
  {Shkolnik}, {Southworth}, {Sozzetti}, \& {Su{\'a}rez
  Mascare{\~n}o}}]{maldo15}
{Maldonado}, J., {Affer}, L., {Micela}, G., {et~al.} 2015, \aap, 577, A132

\bibitem[{{Owen} \& {Wu}(2017)}]{owenwu17}
{Owen}, J.~E. \& {Wu}, Y. 2017, \apj, 847, 29

\bibitem[{Pepe {et~al.}(2014)Pepe, Molaro, Cristiani, Rebolo, Santos, Dekker,
  Mégevand, Zerbi, Cabral, Di~Marcantonio, Abreu, Affolter, Aliverti,
  Allende~Prieto, Amate, Avila, Baldini, Bristow, Broeg, Cirami, Coelho,
  Conconi, Coretti, Cupani, D'Odorico, De~Caprio, Delabre, Dorn, Figueira,
  Fragoso, Galeotta, Genolet, Gomes, González~Hernández, Hughes, Iwert,
  Kerber, Landoni, Lizon, Lovis, Maire, Mannetta, Martins, Monteiro, Oliveira,
  Poretti, Rasilla, Riva, Santana~Tschudi, Santos, Sosnowska, Sousa, Spanó,
  Tenegi, Toso, Vanzella, Viel, \& Zapatero~Osorio}]{pepe14}
Pepe, F., Molaro, P., Cristiani, S., {et~al.} 2014, Astronomische Nachrichten,
  335, 8

\bibitem[{{Perger} {et~al.}(2017){Perger}, {Garc{\'{\i}}a-Piquer}, {Ribas},
  {Morales}, {Affer}, {Micela}, {Damasso}, {Su{\'a}rez-Mascare{\~n}o},
  {Gonz{\'a}lez-Hern{\'a}ndez}, {Rebolo}, {Herrero}, {Rosich}, {Lafarga},
  {Bignamini}, {Sozzetti}, {Claudi}, {Cosentino}, {Molinari}, {Maldonado},
  {Maggio}, {Lanza}, {Poretti}, {Pagano}, {Desidera}, {Gratton}, {Piotto},
  {Bonomo}, {Martinez Fiorenzano}, {Giacobbe}, {Malavolta}, {Nascimbeni},
  {Rainer}, \& {Scandariato}}]{perger17}
{Perger}, M., {Garc{\'{\i}}a-Piquer}, A., {Ribas}, I., {et~al.} 2017, \aap,
  598, A26

\bibitem[{Quarles {et~al.}(2017)Quarles, Quintana, Lopez, Schlieder, \&
  Barclay}]{quarles17}
Quarles, B., Quintana, E.~V., Lopez, E., Schlieder, J.~E., \& Barclay, T. 2017,
  The Astrophysical Journal Letters, 842, L5

\bibitem[{{Quirrenbach} {et~al.}(2016){Quirrenbach}, {Amado}, {Caballero},
  {Mundt}, {Reiners}, {Ribas}, {Seifert}, {Abril}, {Aceituno},
  {Alonso-Floriano}, {Anwand-Heerwart}, {Azzaro}, {Bauer}, {Barrado},
  {Becerril}, {Bejar}, {Benitez}, {Berdinas}, {Brinkm{\"o}ller}, {Cardenas},
  {Casal}, {Claret}, {Colom{\'e}}, {Cortes-Contreras}, {Czesla}, {Doellinger},
  {Dreizler}, {Feiz}, {Fernandez}, {Ferro}, {Fuhrmeister}, {Galadi},
  {Gallardo}, {G{\'a}lvez-Ortiz}, {Garcia-Piquer}, {Garrido}, {Gesa},
  {G{\'o}mez Galera}, {Gonz{\'a}lez Hern{\'a}ndez}, {Gonzalez Peinado},
  {Gr{\"o}zinger}, {Gu{\`a}rdia}, {Guenther}, {de Guindos}, {Hagen}, {Hatzes},
  {Hauschildt}, {Helmling}, {Henning}, {Hermann}, {Hern{\'a}ndez Arabi},
  {Hern{\'a}ndez Casta{\~n}o}, {Hern{\'a}ndez Hernando}, {Herrero}, {Huber},
  {Huber}, {Huke}, {Jeffers}, {de Juan}, {Kaminski}, {Kehr}, {Kim}, {Klein},
  {Kl{\"u}ter}, {K{\"u}rster}, {Lafarga}, {Lara}, {Lamert}, {Laun},
  {Launhardt}, {Lemke}, {Lenzen}, {Llamas}, {Lopez del Fresno},
  {L{\'o}pez-Puertas}, {L{\'o}pez-Santiago}, {Lopez Salas}, {Magan
  Madinabeitia}, {Mall}, {Mandel}, {Mancini}, {Marin Molina}, {Maroto
  Fern{\'a}ndez}, {Mart{\'{\i}}n}, {Mart{\'{\i}}n-Ruiz}, {Marvin}, {Mathar},
  {Mirabet}, {Montes}, {Morales}, {Morales Mu{\~n}oz}, {Nagel}, {Naranjo},
  {Nowak}, {Palle}, {Panduro}, {Passegger}, {Pavlov}, {Pedraz}, {Perez},
  {P{\'e}rez-Medialdea}, {Perger}, {Pluto}, {Ram{\'o}n}, {Rebolo}, {Redondo},
  {Reffert}, {Reinhart}, {Rhode}, {Rix}, {Rodler}, {Rodr{\'{\i}}guez},
  {Rodr{\'{\i}}guez L{\'o}pez}, {Rohloff}, {Rosich}, {Sanchez Carrasco},
  {Sanz-Forcada}, {Sarkis}, {Sarmiento}, {Sch{\"a}fer}, {Schiller}, {Schmidt},
  {Schmitt}, {Sch{\"o}fer}, {Schweitzer}, {Shulyak}, {Solano}, {Stahl},
  {Storz}, {Tabernero}, {Tala}, {Tal-Or}, {Ulbrich}, {Veredas}, {Vico Linares},
  {Vilardell}, {Wagner}, {Winkler}, {Zapatero Osorio}, {Zechmeister},
  {Ammler-von Eiff}, {Anglada-Escud{\'e}}, {del Burgo}, {Garcia-Vargas},
  {Klutsch}, {Lizon}, {Lopez-Morales}, {Ofir}, {P{\'e}rez-Calpena}, {Perryman},
  {S{\'a}nchez-Blanco}, {Strachan}, {St{\"u}rmer}, {Su{\'a}rez}, {Trifonov},
  {Tulloch}, \& {Xu}}]{quirrenbach16}
{Quirrenbach}, A., {Amado}, P.~J., {Caballero}, J.~A., {et~al.} 2016, in
  \procspie, Vol. 9908, Ground-based and Airborne Instrumentation for Astronomy
  VI, 990812

\bibitem[{{Rauer} {et~al.}(2014){Rauer}, {Catala}, {Aerts}, {Appourchaux},
  {Benz}, {Brandeker}, {Christensen-Dalsgaard}, {Deleuil}, {Gizon}, {Goupil},
  {G{\"u}del}, {Janot-Pacheco}, {Mas-Hesse}, {Pagano}, {Piotto}, {Pollacco},
  {Santos}, {Smith}, {Su{\'a}rez}, {Szab{\'o}}, {Udry}, {Adibekyan}, {Alibert},
  {Almenara}, {Amaro-Seoane}, {Eiff}, {Asplund}, {Antonello}, {Barnes},
  {Baudin}, {Belkacem}, {Bergemann}, {Bihain}, {Birch}, {Bonfils}, {Boisse},
  {Bonomo}, {Borsa}, {Brand{\~a}o}, {Brocato}, {Brun}, {Burleigh}, {Burston},
  {Cabrera}, {Cassisi}, {Chaplin}, {Charpinet}, {Chiappini}, {Church},
  {Csizmadia}, {Cunha}, {Damasso}, {Davies}, {Deeg}, {D{\'{\i}}az}, {Dreizler},
  {Dreyer}, {Eggenberger}, {Ehrenreich}, {Eigm{\"u}ller}, {Erikson}, {Farmer},
  {Feltzing}, {de Oliveira Fialho}, {Figueira}, {Forveille}, {Fridlund},
  {Garc{\'{\i}}a}, {Giommi}, {Giuffrida}, {Godolt}, {Gomes da Silva},
  {Granzer}, {Grenfell}, {Grotsch-Noels}, {G{\"u}nther}, {Haswell}, {Hatzes},
  {H{\'e}brard}, {Hekker}, {Helled}, {Heng}, {Jenkins}, {Johansen},
  {Khodachenko}, {Kislyakova}, {Kley}, {Kolb}, {Krivova}, {Kupka}, {Lammer},
  {Lanza}, {Lebreton}, {Magrin}, {Marcos-Arenal}, {Marrese}, {Marques},
  {Martins}, {Mathis}, {Mathur}, {Messina}, {Miglio}, {Montalban}, {Montalto},
  {Monteiro}, {Moradi}, {Moravveji}, {Mordasini}, {Morel}, {Mortier},
  {Nascimbeni}, {Nelson}, {Nielsen}, {Noack}, {Norton}, {Ofir}, {Oshagh},
  {Ouazzani}, {P{\'a}pics}, {Parro}, {Petit}, {Plez}, {Poretti}, {Quirrenbach},
  {Ragazzoni}, {Raimondo}, {Rainer}, {Reese}, {Redmer}, {Reffert},
  {Rojas-Ayala}, {Roxburgh}, {Salmon}, {Santerne}, {Schneider}, {Schou},
  {Schuh}, {Schunker}, {Silva-Valio}, {Silvotti}, {Skillen}, {Snellen}, {Sohl},
  {Sousa}, {Sozzetti}, {Stello}, {Strassmeier}, {{\v S}vanda}, {Szab{\'o}},
  {Tkachenko}, {Valencia}, {Van Grootel}, {Vauclair}, {Ventura}, {Wagner},
  {Walton}, {Weingrill}, {Werner}, {Wheatley}, \& {Zwintz}}]{rauer14}
{Rauer}, H., {Catala}, C., {Aerts}, C., {et~al.} 2014, Experimental Astronomy,
  38, 249

\bibitem[{{Ricker} {et~al.}(2014){Ricker}, {Winn}, {Vanderspek}, {Latham},
  {Bakos}, {Bean}, {Berta-Thompson}, {Brown}, {Buchhave}, {Butler}, {Butler},
  {Chaplin}, {Charbonneau}, {Christensen-Dalsgaard}, {Clampin}, {Deming},
  {Doty}, {De Lee}, {Dressing}, {Dunham}, {Endl}, {Fressin}, {Ge}, {Henning},
  {Holman}, {Howard}, {Ida}, {Jenkins}, {Jernigan}, {Johnson}, {Kaltenegger},
  {Kawai}, {Kjeldsen}, {Laughlin}, {Levine}, {Lin}, {Lissauer}, {MacQueen},
  {Marcy}, {McCullough}, {Morton}, {Narita}, {Paegert}, {Palle}, {Pepe},
  {Pepper}, {Quirrenbach}, {Rinehart}, {Sasselov}, {Sato}, {Seager},
  {Sozzetti}, {Stassun}, {Sullivan}, {Szentgyorgyi}, {Torres}, {Udry}, \&
  {Villasenor}}]{tess14}
{Ricker}, G.~R., {Winn}, J.~N., {Vanderspek}, R., {et~al.} 2014, in \procspie,
  Vol. 9143, Space Telescopes and Instrumentation 2014: Optical, Infrared, and
  Millimeter Wave, 914320

\bibitem[{{Robertson} {et~al.}(2013){Robertson}, {Endl}, {Cochran}, \&
  {Dodson-Robinson}}]{robertson13}
{Robertson}, P., {Endl}, M., {Cochran}, W.~D., \& {Dodson-Robinson}, S.~E.
  2013, \apj, 764, 3

\bibitem[{{Rogers}(2015)}]{rogers15}
{Rogers}, L.~A. 2015, \apj, 801, 41

\bibitem[{Savanov(2012)}]{Savanov2012}
Savanov, I.~S. 2012, Astronomy Reports, 56, 716

\bibitem[{{Selsis} {et~al.}(2007){Selsis}, {Chazelas}, {Bord{\'e}}, {Ollivier},
  {Brachet}, {Decaudin}, {Bouchy}, {Ehrenreich}, {Grie{\ss}meier}, {Lammer},
  {Sotin}, {Grasset}, {Moutou}, {Barge}, {Deleuil}, {Mawet}, {Despois},
  {Kasting}, \& {L{\'e}ger}}]{selsis07}
{Selsis}, F., {Chazelas}, B., {Bord{\'e}}, P., {et~al.} 2007, \icarus, 191, 453

\bibitem[{{Shields} {et~al.}(2016){Shields}, {Ballard}, \&
  {Johnson}}]{shields16}
{Shields}, A.~L., {Ballard}, S., \& {Johnson}, J.~A. 2016, \physrep, 663, 1

\bibitem[{{Sinukoff} {et~al.}(2016){Sinukoff}, {Howard}, {Petigura},
  {Schlieder}, {Crossfield}, {Ciardi}, {Fulton}, {Isaacson}, {Aller},
  {Baranec}, {Beichman}, {Hansen}, {Knutson}, {Law}, {Liu}, {Riddle}, \&
  {Dressing}}]{sinukoff16}
{Sinukoff}, E., {Howard}, A.~W., {Petigura}, E.~A., {et~al.} 2016, \apj, 827,
  78

\bibitem[{{Sozzetti} {et~al.}(2013){Sozzetti}, {Bernagozzi}, {Bertolini},
  {Calcidese}, {Carbognani}, {Cenadelli}, {Christille}, {Damasso}, {Giacobbe},
  {Lanteri}, {Lattanzi}, \& {Smart}}]{sozzetti13}
{Sozzetti}, A., {Bernagozzi}, A., {Bertolini}, E., {et~al.} 2013, in European
  Physical Journal Web of Conferences, Vol.~47, European Physical Journal Web
  of Conferences, 03006

\bibitem[{{Tuomi} {et~al.}(2014){Tuomi}, {Jones}, {Barnes},
  {Anglada-Escud{\'e}}, \& {Jenkins}}]{tuomi14}
{Tuomi}, M., {Jones}, H.~R.~A., {Barnes}, J.~R., {Anglada-Escud{\'e}}, G., \&
  {Jenkins}, J.~S. 2014, \mnras, 441, 1545

\bibitem[{{Van Eylen} {et~al.}(2017){Van Eylen}, {Agentoft}, {Lundkvist},
  {Kjeldsen}, {Owen}, {Fulton}, {Petigura}, \& {Snellen}}]{vaneylen17}
{Van Eylen}, V., {Agentoft}, C., {Lundkvist}, M.~S., {et~al.} 2017, ArXiv
  e-prints [\eprint[arXiv]{1710.05398}]

\bibitem[{{Vanderburg} \& {Johnson}(2014)}]{vanderjohns14}
{Vanderburg}, A. \& {Johnson}, J.~A. 2014, \pasp, 126, 948

\bibitem[{{Vanderburg} {et~al.}(2016){Vanderburg}, {Plavchan}, {Johnson},
  {Ciardi}, {Swift}, \& {Kane}}]{vander16}
{Vanderburg}, A., {Plavchan}, P., {Johnson}, J.~A., {et~al.} 2016, \mnras, 459,
  3565

\bibitem[{{Zechmeister} \& {K{\"u}rster}(2009)}]{zech09}
{Zechmeister}, M. \& {K{\"u}rster}, M. 2009, \aap, 496, 577

\bibitem[{{Zeng} {et~al.}(2017a){Zeng}, {Jacobsen}, {Hyung}, {Vanderburg},
  {Lopez-Morales}, {Sasselov}, {Perez-Mercader}, {Petaev}, {Latham}, {Haywood},
  \& {Mattson}}]{zeng17b}
{Zeng}, L., {Jacobsen}, S.~B., {Hyung}, E., {et~al.} 2017a, in Lunar and
  Planetary Inst.~Technical Report, Vol.~48, Lunar and Planetary Science
  Conference, 1576

\bibitem[{Zeng {et~al.}(2017b)Zeng, Jacobsen, \& Sasselov}]{zeng17c}
Zeng, L., Jacobsen, S.~B., \& Sasselov, D.~D. 2017b, Research Notes of the AAS,
  1, 32

\bibitem[{{Zeng} \& {Sasselov}(2013)}]{zeng13}
{Zeng}, L. \& {Sasselov}, D. 2013, \pasp, 125, 227

\bibitem[{{Zeng} {et~al.}(2016){Zeng}, {Sasselov}, \& {Jacobsen}}]{zeng16}
{Zeng}, L., {Sasselov}, D.~D., \& {Jacobsen}, S.~B. 2016, \apj, 819, 127

\end{thebibliography}
\end{document}